\documentclass{article}
\usepackage{latexsym,amssymb,amsmath, a4wide, theorem, graphicx}
\usepackage{appendix}

\font\twrm=cmr10
 \newcommand{\nc}{\newcommand}
 \nc{\di}{\displaystyle} \nc{\ra}{\rightarrow}
 \nc{\al}{\alpha} \nc{\ve}{\varepsilon} \nc{\vp}{\varphi}
 \nc{\Ga}{\Gamma} \nc{\ga}{\gamma} \nc{\Om}{\Omega}
 \nc{\om}{\omega} \nc{\la}{\lambda} \nc{\Si}{\Sigma}
 \nc{\si}{\sigma} \nc{\Da}{\Delta} \nc{\da}{\delta}
 \nc{\na}{\nabla} \nc{\pa}{\partial} \nc{\ti}{\times}
 \nc{\N}{{\mathbb N}} \nc{\Z}{{\mathbb Z}} \nc{\C}{{\mathbb C}}
 \nc{\R}{{\mathbb R}} \nc{\V}{{\cal V}} \nc{\ek}{\protect\\[1ex]}
 \nc{\eq}[1]{\mbox{\twrm {(\ref{E#1})}}}
 \nc{\ome}{\Om^\ve} \nc{\les}{\lesssim}
  \nc{\lan}{\langle} \nc{\ran}{\rangle}
 \nc{\PP}{(\mathbb{BL})^i_{\la,\tx'}}
 \nc{\tx}{\texttt{x}} \nc{\ty}{\texttt{y}}
 \nc{\hty}{\hat{\ty}} \nc{\hb}{\hat{\beta}}
 \nc{\curl}{\text{curl}} \nc{\tbeta}{\tilde{\beta}}
 \nc{\tom}{\tilde{\om}} \nc{\ttp}{\tilde{\tilde{p}}}
 \nc{\ttvp}{\tilde{\tilde{\vp}}} \nc{\ha}{\frac{1}{2}} \nc{\s}{\tilde}
\renewcommand{\div}{{\rm div}\,}
\nc{\supp}{{\rm supp}\,}

\newtheorem{lem}{Lemma}[section]
\newtheorem{theo}[lem]{Theorem}

\newtheorem{rem}[lem]{Remark}

\newtheorem{tdefi}[lem]{Definition}
\numberwithin{equation}{section}

\title{Effective wall-laws for the Stokes equations \\over curved rough boundaries}

\author{Myong-Hwan Ri\\
\small {Institute of Mathematics, State Academy of Sciences, DPR
Korea}
}
\date{\today}

\begin{document}

\maketitle
%

\begin{abstract}
We derive effective wall-laws for Stokes systems with inhomogeneous
boundary conditions in three dimensional bounded domains with curved rough
boundaries. No-slip boundary condition is given on the locally
periodic rough boundary parts with characteristic roughness size
$\ve$ and boundary data is assumed to be supported in the
nonoscillatory smooth boundary.

Based on the analysis of a boundary layer cell problem depending on
 geometry of the fictitious boundary and roughness shape, boundary layer
approximations are constructed using orthogonal tangential vectors
and normal vector on the fictitious boundary, which have
$O(\ve^{3/2})$-order in $L^2$-norm and
$O(\ve)$-order in energy-norm. Then, a Navier wall-law with error
estimates of $O(\ve^{3/2})$-order in $L^2$-norm and $O(\ve)$-order
in $W^{1,1}$-norm is obtained, which is proved to be irrespective of
the choice of the orthogonal tangent vectors.

\end{abstract}


\section{Introduction}

Rough boundary problems have many practical applications
in aerodynamics, electromagnetism, hydrodynamics and hemodynamics,
etc. Direct numerical computation around rough boundaries is usually
out of reach for the time being since the problems have both
macroscopic and microscopic scales and hence need lots of computational
burden. Therefore one usually changes the boundary condition on
rough boundary with a new boundary condition on a regularized
fictitious boundary close to the rough boundary, that is so-called a
{\it wall-law}. For viscous fluid flows, no-slip boundary condition
at the rough wall is replaced by a type of Navier slip boundary
condition, {\it Navier wall-law}, at the fictitious boundary. The
derivation of Navier wall-laws are also important for shape optimization of
roughness for better drag reduction since the  procedure of
shape optimization for drag reduction requires to know
{\it a priori} the Navier's coefficient in the slip boundary
condition.
\par
In this article we study effective wall-laws for the Stokes system
\begin{equation}
\label{E3.1}
\begin{array}{rcll}
 -\Da u^\ve + \na p^\ve & =& f &\quad\text{in }\ome,\ek
 \div u^\ve & = &0  &\quad \text{in }\ome, \ek
 u^\ve & = & \psi &\quad  \text{in }\pa\ome,
\end{array}
\end{equation}
where $\ome\subset\R^3$ is a bounded and simply connected domain and
its sufficiently smooth boundary $\pa\ome$ consists of a
nonoscillatory part and a rough part formed by locally periodic
microscopic rugosities of characteristic size $O(\ve)$. Boundary
data is assumed to be supported in the nonoscillatory part of
$\pa\ome$.
 \par  There is a number of papers dealing with effective wall-laws
 for Stokes and Navier-Stokes equations, see e.g.
\cite{APV98, AcTaVaPi98, Al92, AmClCaSi01, AmBoMaGa10, BaTaVa02,
JM00, JM01, JMN02, JM03, MoPiVa98}
   and the references therein for
the case of periodic roughness and \cite{BaGe08, Ge09} for the case
of nonperiodic random roughness.
 Moreover, one can find results concerning
   explicit or implicit wall-laws for Poisson equations,
  see, e.g. \cite{AP95, AlAm99, AmBoMaGa04, BoBrMi08, BoBrMi08-2, BrMi10,
BrMi06, BrMi08-2, MaVa02, MaVa06, NNM06}. Here, we note that the
Poisson equations describe simplified flows which are uniform in
longitudinal direction. The main techniques used to derive effective
wall-laws are domain decomposition and multiscale asymptotic
expansions.
\par Most rough boundaries we meet in reality are curved
boundaries, and for practical applications, results of flat rough
boundaries may be applied to curved rough boundary problems with
small curvature to some extent. However,  if the curvature is
considerably large, for more accurate analysis near the rough
surfaces and for determination of micro-roughness shape giving
better performance of drag reduction,  the curved rough boundary
must be considered as it is. We note that most of above mentioned
references concern the flat rough boundaries, while for references
dealing with wall-laws for curved rough boundaries, we refer to e.g.
\cite{AP95, MaVa02, MaVa06, NNM06}.
 In a pioneering work \cite{AP95} a first order wall-law for the Poisson equation
 in a ring with many small holes near the outer boundary
 was obtained using domain decomposition techniques. Later,
in \cite{MaVa02, MaVa06}, first and second order wall-laws for
Poisson equations in general two-dimensional annular domains with
curved rough boundaries were obtained by combining techniques of domain
decomposition and two-scale asymptotic expansions. We note that
two-dimensional problems correspond to the case where longitudinal
grooves form rough surfaces.
\par
   Wall-laws for multi-dimensional Poisson
problem over curved rough boundary were obtained in \cite{NNM06}.
More precisely, for Poisson problem with homogeneous Dirichlet
boundary condition on curved compact boundary with locally periodic
roughness on it the authors constructed suitable approximations of
$O(\ve^{3/2})$-order in $L^2$-norm and $O(\ve)$-order in energy norm
based on analysis of a boundary layer cell problem.
 Then, a wall-law with the same order of error estimates as the approximations in
interior domains was derived.

 We refer to a review article \cite{Mi09}
 for more details  on the derivation and analysis of wall laws of fluid flows.

\par
Motivated by \cite{NNM06}, in this article we address derivation of
wall-laws for {\it inhomogeneous boundary
 value problem for the Stokes systems \eq{3.1}}
 over curved rough boundaries. We note that the system \eq{3.1} may be used to analyze
exterior fluid flows, if the boundary has two components and boundary
data is supported only in outer non-oscillatory
boundary part. Furthermore, if
 the rough boundary part and non-oscillatory boundary part are adjacent,
  the system \eq{3.1}
 may be used for local analysis of fluid flows near a curved rough
 surface that can be a part of boundary of any type of objects.

To achieve our goal, first, we analyse a
boundary layer cell problem depending on the geometry of the
fictitious boundary and roughness shape, that is elliptic in the sense of Agmon, Douglis and
Nirenberg, see $\PP$ in subsection 3.2, by using technique of
Fourier series expansion. Then we construct boundary
layer approximations of $O(\ve^{3/2})$-order in $L^2$-norm and
$O(\ve)$-order in energy norm using the orthogonal tangential
vectors and normal vector on the fictitious surface. Using these approximations we obtain
an effective Navier wall-law which is shown to be irrespective of the
choice of the orthogonal tangential vector fields and has error of $O(\ve^{3/2})$-order in
$L^2$-norm and $O(\ve)$-order in $W^{1,1}$-norm. The main theorems of the paper are Theorems
 \ref{T3.14}, \ref{T3.15} and \ref{T3.20}.
\par
 Dealing with the Stokes system, we are encountered with
  additional difficulties compared to the Poisson problem,
which are mainly related to the difference between vectorial and scalar
case as well as the structural complicatedness of the
Stokes system over Poisson equation.
The main difference from the case of scalar
Poisson equation is that for the construction of the approximations
and wall-laws we need to consider curvilinear systems of tangential
vectors and normal vector on the fictitious surface;
we should take careful observations of dependence of approximations
on the local curvilinear system.
Moreover,  due to inhomogeneous boundary condition,
we need some sophisticated techniques  using cut-off functions
in construction and estimates of boundary layer
approximations so that artificial vertical layer flows around the nonoscillatory
 boundary part could not be generated thus ensuring the required
approximation order near the edge between nonoscillatory
and oscillating parts of boundary.
\par
For simplicity we consider the case of spacial dimension $n=3$, but
 the result of the paper can be directly extended to the case of
  $n>3$ without essential change.

  This paper is organized as follows. In Section 2 we describe the rough
  domain considered in the paper and give the main notations.
  Section 3, the major part of the paper, consists of several subsections.
  Estimates of  Dirichlet wall-law are given in subsection 3.1 and a boundary
  layer cell problem is analyzed in subsection 3.2. Subsections 3.3 and
  3.4 concern the construction of local and global
  boundary layer correctors, respectively.
  In subsection 3.5 boundary layer approximations are
  constructed and an effective Navier wall-law
  with higher order error estimates is derived. Finally,
  in Appendix we give a refined analysis
  for divergence problem ensuring the estimate constant for a
  solution of divergence equation in our rough domain
  being independent of micro-roughness size $\ve$.

\section{Domains with rough boundaries and main notations}\label{notation}
 We give description on the rough domain and notations.
 The domain $\ome\subset\R^3$ is bounded with its boundary
$\pa\ome$ consisting of rough part $\Ga_0$ and nonoscillatory smooth part
$\Ga_1$, i.e.,
$$\pa\ome=\Ga_0\cup\Ga_1,\quad \Ga_0\cap\Ga_1=\emptyset,$$
where $\Ga_1$ is closed and $\Ga_0$ consists of finite locally
$\ve$-periodic oscillating parts with microscopic size $\ve$.
The domain $\ome$ is divided into $\Om$ and $\ome\setminus\Om$ by an
open and nonoscillatory sufficiently smooth surface $\Ga$ (fictitious boundary)
such that
$\Ga$ is at the distance of $O(\ve)$ from $\Ga_0$ and
$\pa\Om=\Ga\cup\Ga_1$, $\Ga\cap\Ga_1=\emptyset$. If $\Ga$ and $\Ga_1$ are
adjacent, we assume $\pa\Om\in C^{0,1}$.

Denoting by $\nu(x')$ the outward normal vector for $\Om$ at
$x'\in\Ga$, suppose that there is some positive $\da=O(1)$ such that
the mapping
$$\Upsilon: \Ga\times(-\da,\da)\rightarrow \R^3,
\Upsilon(x',t)=x'+t\nu(x')$$ is diffeomorphism. Moreover, suppose
that there are some bounded open sets $U_i, i=1,\ldots, N,$ of
$\R^2$, $V_i, i=1,\ldots, N,$ of $\Ga$ and diffeomorphism
$\varphi_i: U_i\ra V_i, i=1,\ldots, N,$ such that
$\{\varphi_i,U_i,V_i\}_{i=1}^N$ is a chart of $\Ga$ and the rough
surface $\Ga_0$ is expressed by
$$\begin{array}{l}
\Ga_0= \{\Upsilon(x',\ga^\ve(x')): x'\in\Ga\},\\
\ga^\ve(x')=\ve\ga_i\big(\varphi_i^{-1}(x'),
\frac{\varphi_i^{-1}(x')}{\ve}\big), x'\in V_i,i=1,\ldots,N,
 \end{array}$$
  where $\ga_i\geq 0$ defined in $U_i\times \R^2$
 is $(1,1)$-periodic with respect to the second variable
 and may take multi-values.
  In this sense the rough boundary part $\Ga_0$ is locally $\ve$-periodic.
 In addition, let
 \begin{equation}
 \label{E1.3}
|\ga^\ve(x')|\leq \ve M< \frac{\da}{2}, x'\in \Ga,
 \end{equation}
and put
$$\begin{array}{l}
\Ga_\da:=\Upsilon(\Ga\times(-\da,\da)),\,\,\,\Ga_\da^\ve=\Ga_\da\cap\ome,\\
\Ga_{\da,i}:=\Upsilon(V_i\times(-\da,\da)),\,\,\,\Ga_{\da,i}^\ve=
\Ga_{\da,i}\cap\ome, i=1,\ldots,N.
 \end{array}$$
 Obviously,
$\Ga_\da^\ve=\cup_{i=1}^N\Ga_{\da,i}^\ve$,
$\Ga_\da^\ve\supset\Om^\ve\setminus\Om$.

It is natural to assume that $\Om^\ve$ can be expressed as a type of domain
\begin{equation}
\label{E1.2}
\Om^\ve=\bigcup_{j=1}^M G^{(j)\ve},\quad
  G^{(j)\ve}=G_0^{(j)}\cup \bigcup_{k=1}^{m_j}G_k^{(j)},\quad j=1,\ldots,M,
\end{equation}
where  $M\sim O(1),|G_0^{(j)}| \sim O(1)$, $|G_k^{(j)}|\sim O(\ve^3)$ and $m_j\sim O(\frac{1}{\ve^2})$,
 $k=1,\ldots, m_j$, $j=1,\ldots,M$,
and
$$G_0^{(j)}\cap G_k^{(j)}\neq \emptyset, G_k^{(j)}\cap G_l^{(j)}=\emptyset,
k\neq l, 1\leq k,l \leq m_j,$$
for each $j\in \{1,\ldots,M\}$.
We assume further that
each $G_k^{(j)}$, $k=0,\ldots, m_j, j=1,\ldots,M$,
is a star-shaped domain with respect to some ball of radius $R_k^{(j)}$
and $\frac{\da(G_k^{(j)})}{R_k^{(j)}}\sim O(1)$, where $\da(G_k^{(j)})$ is
the diameter of $G_k^{(j)}$.
These additional assumptions will guarantee that the estimate constant for a solution
of divergence equation is independent of $\ve$, see Appendix.

\par
For consideration of inhomogeneous boundary condition
we use the notation for $\Ga'$ and $\Om_0$ as
 $\Ga'\equiv \Ga$ and $\Om_0\equiv \Om$ if $\Ga$ and $\Ga_1$ are
 components of $\pa\Om$, and, if $\Ga$ and $\Ga_1$ are adjacent,
  $$\Ga'\equiv\{\tx'\in\Ga: d(\tx',\pa\Ga)\geq\ve\}$$
  and $\Om_0$ is a sufficiently smooth domain satisfying
$$\Om\subset\Om_0\subset\Om^\ve, \quad\Ga_1\cup\Ga'\subset\pa\Om_0
\text{  and } \Om_0\setminus\Om\text{  has thickness of size }
O(\ve^2).$$
 Then, obviously, $|\Om_0\setminus\Om|\leq O(\ve^3)$.
 Let
$\Ga'_\da\equiv\Upsilon(\Ga'\ti(-\da,\da))$.
\begin{figure}
\label{roughdomains}
\begin{center}
\includegraphics[scale=0.5]{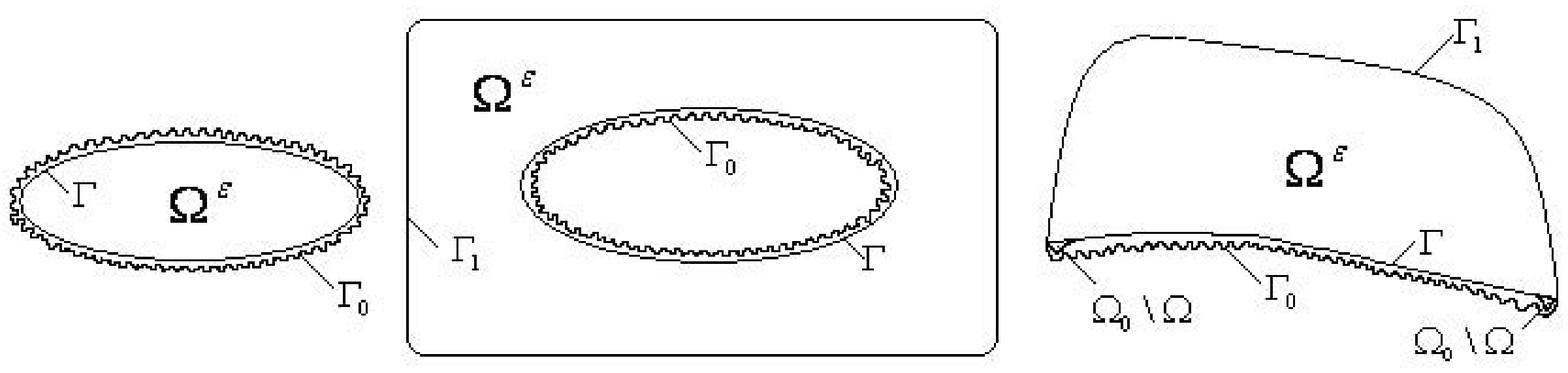}

\end{center}
\caption{Rough domains}
\end{figure}
\par
As usual, $\N$ is the set of all natural numbers,
$\N_0=\N\cup\{0\}$, and $\Z$ is the set of all integers. For a
domain $G\subset\R^n$ its closure is denoted by $\bar{G}$ and its
boundary by $\pa G$. We do not distinguish between
spaces of scalar- and vector-, or even tensor-valued functions
 as long as no confusion arises. For Lebesgue, Sobolev spaces on a domain or
boundary we use standard notations $L^r$, $W^{k,r}, W^{k,r}_0$,
$1\leq r \leq\infty$, $k\in \Z$, respectively.
We use notation $L^2_{(m)}(G):=\{\varphi\in L^2(G):
\int_{G}\varphi(x)\,dx=0\}$. Let
$H^1\equiv W^{1,2}, H^1_0\equiv W^{1,2}_0$ and $H^{-1}$ the dual of
$H^1_0$. The closures in $H^{1}(G)$ and $L^r(G)$ of the set $\{u\in
C^{\infty}_0(G): \div u=0\}$ are denoted by
  $H^1_{0,\si}(G)$ and $L^r_{\si}(G)$, respectively.
The notation $A\les B$ ($A\gtrsim B$) implies $A\leq cB$ ($A\geq
cB$) with constant $C$ independent of $\ve$.


\section{Effective wall-laws for the Stokes system}
\par
Let us assume for the data of \eq{3.1} that
 \begin{equation}
 \label{E2.1n}
f\in L^q(\ome), \psi\in W^{2-1/q,q}(\pa\ome), q\geq 2,\quad\supp
\psi\subset \Ga_1, \int_{\Ga_1} \psi\cdot\nu dx=0.
 \end{equation}
 It is well known, cf. e.g. \cite{Ga94-1}, that the system \eq{3.1} has
a unique solution $\{u^\ve, p^\ve\}$ satisfying
$u^\ve \in W^{2,q}(\ome),\, p^\ve \in W^{1,q}(\ome)$
and
\begin{equation}
\label{E3.2}
 \|u^\ve\|_{W^{2,q}(\Om^\ve)}+\|p\|_{W^{1,q}(\Om^\ve)}\leq
C(\Om^\ve)(\|f\|_{L^q(\Om^\ve)}+\|\psi\|_{W^{2-1/q,q}(\Ga_1)}).
\end{equation}
\subsection{Dirichlet wall-law}
 Consider the approximation of the system \eq{3.1} as
 \begin{equation}
 \label{E3.3}
 \begin{array}{rcll}
 -\Da u + \na p & =& f &\quad\text{in }\Om_0,\\
 \div u & = &0  &\quad \text{in }\Om_0,\\
 u & = & \psi &\quad  \text{in }\Ga_1,\\
u & = & 0 &\quad  \text{in }\pa\Om_0\setminus\Ga_1.
\end{array}
\end{equation}
The system \eq{3.3} has a unique solution $\{u,p\}\in  W^{2,q}(\Om_0)\ti W^{1,q}(\Om_0)$
such that
\begin{equation}
\label{E3.4}
\begin{array}{c}
 \|u\|_{W^{2,q}(\Om_0)}+\|p\|_{W^{1,q}(\Om_0)}\leq
C(\Om_0)(\|f\|_{L^q(\Om_0)}+\|\psi\|_{W^{2-1/q,q}(\Ga_1)}).
\end{array}
\end{equation}
\par
\begin{lem}
{\rm \label{L2.3}
  If $\varphi\in H^1(\ome\setminus\Om)$,
$\vp|_{\Ga_0}=0$, then
\begin{equation}
 \label{E2.5n}
 \|\varphi\|_{L^2(\pa\Om_0\setminus\Ga_1)}+\|\varphi\|_{L^2(\Ga)}\les
\ve^\ha\|\na\varphi\|_{L^2(\ome\setminus\Om)}.
 \end{equation}
 In addition, if $\varphi\in H^2(\ome)$, then
 \begin{equation}
 \label{E3.8nn}
 \|\varphi\|_{L^2(\pa\Om_0\setminus\Ga_1)}+\|\vp\|_{L^2(\Ga)} \les \ve \|\vp\|_{H^2(\ome)}.
 \end{equation}
 }
 \end{lem}
 \textbf{Proof.}
For $x'\in\Ga$ let $l(x')$ and $l_0(x')$ be the distances from $x'$
to the intersection points of the outer normal line  at $x'$ for
$\Om$ with $\pa\Om_0\setminus\Ga_1$ and $\Ga_0$, respectively. Let
$\tilde{\vp}$ be extension of $\vp$ by $0$ to
$\Ga_\da\setminus\Ga_\da^\ve$.
 Then we get by $\vp|_{\Ga_0}=0$ and Minkowski's inequality that
\begin{equation}
 \label{E3.7n}
\begin{array}{rcl}
 \|\varphi\|_{L^2(\pa\Om_0\setminus\Ga_1)}
 &=&\Big(\int_\Ga|\int_{l_0(x')}^{l(x')}\frac{\pa\vp}{\pa\tx_3}(x',\tx_3)\,d\tx_3|^2\,dx'\Big)^{1/2}\ek
 &\leq&\Big(\int_\Ga\big(\int_0^{M\ve}|\frac{\pa\tilde{\vp}}{\pa\tx_3}(x',\tx_3)|\,d\tx_3\big)^2\,dx'\Big)^{1/2}\ek
&\les&
\int_0^{M\ve}\Big(\int_\Ga|\frac{\pa\tilde{\vp}}{\pa\tx_3}(x',\tx_3)|^2\,dx'\Big)^{1/2}\,d\tx_3.
\end{array}
 \end{equation}
Then, by H\"older inequality one gets
$$\|\varphi\|_{L^2(\pa\Om_0\setminus\Ga_1)}
\les
\ve^{1/2}\|\frac{\pa\tilde{\vp}}{\pa\tx_3}(x',\tx_3)\|_{L^2(0,M\ve;L^2(\Ga))}
 \les \ve^{1/2} \|\na\vp\|_{L^2(\ome\setminus\Om)}.$$
 The estimate of $\|\varphi\|_{L^2(\Ga)}$ can be obtained in the
same way. Thus, \eq{2.5n} is proved.
\par
Let us prove \eq{3.8nn}. Let $\ttvp\in H^2(\ome\cup\Ga_\da)$ be an
extension of $\vp$ satisfying
\begin{equation}
\label{E3.2nn}
 \|\frac{\pa \vp}{\pa\tx_3}\|_{H^1(\Ga_\da)}\leq C \|\frac{\pa \vp}{\pa\tx_3}\|_{H^1(\ome)}
 \end{equation}
  with constant $C>0$ independent of $\ve$. The existence of such an extension
  is guaranteed by Sobolev extension theorem.
  Note that, due to the continuous embedding
$$H^1((-\da,M\ve); L^2(\Ga))\hookrightarrow
L^\infty((-\da,M\ve);L^2(\Ga))$$ with an embedding constant
independent of $\ve$ and \eq{3.2nn}, one has
$$\big\|\frac{\pa\tilde{\vp}}{\pa\tx_3}\big\|_{L^\infty(0,M\ve;
L^2(\Ga))}\les \big\|\frac{\pa\ttvp}{\pa\tx_3}\big\|_{H^1(-\da,M\ve;
L^2(\Ga))}\les \|\vp\|_{H^2(\ome)}.$$
Consequently, we can proceed in \eq{3.7n} as
$$\int_0^{M\ve}\Big(\int_\Ga|\frac{\pa\tilde{\vp}}{\pa\tx_3}(x',\tx_3)|^2\,dx'\Big)^{1/2}\,d\tx_3\ek
\les
\int_0^{M\ve}\|\frac{\pa\ttvp}{\pa\tx_3}(\cdot,\tx_3)\|_{L^2(\Ga)}\,d\tx_3
\les \ve \|\vp\|_{H^2(\ome)}.$$ Combining this inequality with
\eq{3.7n} yields the required estimate for
$\|\varphi\|_{L^2(\pa\Om_0\setminus\Ga_1)}$ in \eq{3.8nn}. The
estimate of $\|\varphi\|_{L^2(\Ga)}$ in \eq{3.8nn} can be obtained
in the same way.

Thus, the proof of the lemma is complete. \vspace*{0.3cm}\hfill $\Box$
 \par
We have the following theorem on the error estimate for the
zeroth order approximation system \eq{3.3}.
\begin{theo}
\label{T3.1} {\rm Let $u^\ve$ be the solution to \eq{3.1} and let
$\s{u},\s{p}$ be extensions of $u,p$ by $0$ to $\ome$, respectively.
 Then,
$$\|\na(u^\ve-\s{u})\|_{L^2(\ome)}\les\ve^{1/2}(\|f\|_{L^2(\ome)}+\|\psi\|_{H^{3/2}(\Ga_1)}),$$
$$\|u^\ve-\s{u}\|_{L^2(\ome)}\les \ve(\|f\|_{L^2(\ome)}+\|\psi\|_{H^{3/2}(\Ga_1)}).$$
}
\end{theo}
{\bf Proof.} Let $v=u^\ve-\s{u}$ and $s=p^\ve-\s{p}$. Then,
obviously, $v\in H^1_{0,\si}(\ome)$ and for any $\vp\in
H^1_{0,\si}(\ome)$
 we have using integration by
parts that
 \begin{equation}
 \label{E3.6n}
(\na v,\na\vp)_{L^2(\ome)}=(-\Da v+\na s,\vp)_{\ome}
 =(f,\vp)_{\ome\setminus\Om_0}-\int_{\pa\Om_0\setminus\Ga_1}(\frac{\pa
u}{\pa \nu}-p\nu)\cdot\vp\,dx,
 \end{equation}
where $(\cdot, \cdot)$ denotes either $L^2$-scalar product or
duality paring between $H^{-1}$ and $H^{1}_0$.
By Poincar\'e's inequality one has
 \begin{equation}
 \label{E3.4n}
 |(f,\vp)_{\ome\setminus\Om_0}|\leq
\|f\|_{L^2(\ome\setminus\Om_0)}\|\vp\|_{L^2(\ome\setminus\Om_0)}\les
\ve\|f\|_{L^2(\ome\setminus\Om_0)}\|\na\vp\|_{L^2(\ome\setminus\Om_0)}
 \end{equation}
and, by Lemma \ref{L2.3} and \eq{3.4},
 $$\begin{array}{rcl}
 |\int_{\pa\Om_0\setminus\Ga_1}(\frac{\pa u}{\pa
\nu}-p\nu)\cdot\vp\,dx|
 &\leq&
 \|\frac{\pa u}{\pa\nu}-p\nu\|_{L^2(\pa\Om_0\setminus\Ga_1)}\|\vp\|_{L^2(\pa\Om_0\setminus\Ga_1)}\ek
 &\les&
 \ve^{1/2}(\|u\|_{H^2(\Om_0)}+\|p\|_{H^1(\Om_0)})\|\na\vp\|_{L^2(\ome\setminus\Om)}\ek
&\les& \ve^{1/2}(\|f\|_{L^2(\ome)}
+\|\psi\|_{H^{3/2}(\Ga_1)})\|\na\vp\|_{L^2(\ome\setminus\Om)}.
\end{array}
$$
Therefore, the first inequality of the theorem is proved.
\par
 Let us prove the second inequality of the theorem.
Let $A$ be the Stokes operator in $L^2_\si(\ome)$, i.e.,
$$D(A)=H^2(\ome)\cap H^1_{0,\si}(\ome),\quad A\vp:=-P\Da \vp,$$
 where $P$ is the Helmholtz projection of $L^2(\ome)$ onto $L^2_\si(\ome)$.
 It is well known that
 \begin{equation}
  \label{E3.3n}
\|\vp\|_{H^2(\ome)}\les \|A \vp\|_{L^2_\si(\ome)}\les
\|\vp\|_{H^2(\ome)},\quad \forall \vp\in D(A),
\end{equation}
cf. \cite{Te77}.

 Fix any $\vp\in D(A)$. Then, from \eq{3.6n} one has
 $$\begin{array}{rcl}
(v, A\vp)_{L^2_\si(\ome)}=(\na v,
\na\vp)_{L^2(\ome)}=(f,\vp)_{\ome\setminus\Om_0}-\int_{\pa\Om_0\setminus\Ga_1}(\frac{\pa
u}{\pa \nu}-p\nu)\cdot\vp\,dx.
\end{array}
 $$
 Note that, by Lemma \ref{L2.3}, (ii) and \eq{3.4},
   $$\begin{array}{rcl}
 |\int_{\pa\Om_0\setminus\Ga_1}(\frac{\pa u}{\pa
\nu}-p\nu)\cdot\vp\,dx|
 &\leq&
 \|\frac{\pa u}{\pa\nu}-p\nu\|_{L^2(\pa\Om_0\setminus\Ga_1)}\|\vp\|_{L^2(\pa\Om_0\setminus\Ga_1)}\ek
 &\les& \ve(\|f\|_{L^2(\ome)}
+\|\psi\|_{H^{3/2}(\Ga_1)})\|\vp\|_{H^2(\ome)}.
\end{array}
$$
Therefore, in view of \eq{3.4n}, \eq{3.3n}, we get
$$|(v, A\vp)_{L^2_\si(\ome)}|\les \ve(\|f\|_{L^2(\ome)}
+\|\psi\|_{H^{3/2}(\Ga_1)})\|A\vp\|_{L^2_\si(\ome)},$$
which yields the second inequality of the theorem
since the range of $A$ is $L^2_\si(\ome)$.

The proof of the theorem is complete.\hfill $\Box$

\subsection{Boundary layer analysis}
In order to derive a wall-law of higher order approximation for
\eq{3.1} we analyze the boundary layer near the rough boundary.
\par
 For $x\in \Ga_{\da,i}$ let
$$x=\Phi_i(\tx):=\Upsilon(\vp_i(\tx'),\tx_3)=\vp_i(\tx')+\tx_3\nu(\vp_i(\tx')),
\,\,\tx=(\tx',\tx_3)\in U_i\ti (-\da,\da),$$ (see Section 2 for
$\vp_i$).

Based on  the expression of gradient $\na_x$, divergence
$\text{div}_x$ and Laplacian $\Da_x$ with respect to the coordinate
$\tx$, that is,
\begin{equation}
\label{E3.12n}
 \begin{array}{l}
 \na_xg=(D_x \Phi_i^{-1})^T(\na_\tx g(\vp_i(\tx'))\circ
 \Phi_i^{-1})=(D_\tx \Phi_i)^{-T}(\na_\tx g(\vp_i(\tx'))\circ
 \Phi_i^{-1}),\ek
\text{div}_x h= \text{div}_\tx \big((D_\tx \Phi_i)^{-1}
h(\vp_i(\tx'))\big)\circ
 \Phi_i^{-1},\ek
 \Da_x g =\text{div}_\tx \big((D_\tx \Phi_i)^{-1}
(D_\tx \Phi_i)^{-T}\na_\tx g(\vp_i(\tx'))\big)\circ
 \Phi_i^{-1},
 \end{array}
 \end{equation}
where $D_\tx \Phi_i$ is Jacobian matrix for $\Phi$, we introduce
matrices $A_i(\tx), B_i(\tx)$ as
 \begin{equation}
 \label{E3.0}
 \begin{array}{l}
  B_i(\tx'):=(D_\tx \Phi_i(\tx',0))^{-T}=(D_{\tx'}\varphi_i(\tx'),\nu(\vp_i(\tx'))^{-T},\quad
  A_i(\tx'):= B_i(\tx')^TB_i(\tx').
  \end{array}
  \end{equation}
Note that
$$A_i(\tx')=
\left(\begin{array}{cc}
                            (D_{\tx'}\varphi_i(\tx')^T D_{\tx'}\varphi_i(\tx'))^{-1}& 0\\
                            0                & 1
                            \end{array}
                      \right),\quad \tx'\in U_i.$$
Then we formulate the boundary layer cell problem $\PP$ with
parameter $\tx'\in U_i$ and $\la \in\R^3$:
$$
\PP: \hspace{0.1cm}\left\{
  \begin{array}{cl}
  -\div_{\ty}\big( A_i(\tx')\na_\ty\beta(\tx',\ty)\big)+B_i(\tx')\na_\ty\om(x',\ty)=0,
         &\text{in }Z_{BL}\setminus S,\ek
     \div_{\ty}\big( B_i(\tx')^T\beta(\tx',\ty)\big)=0,& \text{in }Z_{BL}\setminus S,\ek
     [\beta(\tx',\ty)]_S=0,&\ek
     \big[\frac{\pa\beta}{\pa\ty_3}-\om\nu(\vp_i(\tx'))\big]_S=\la,&\ek
     \beta(\tx',\ty)=0,&\text{on }\{\ty_3=\ga_i(\tx',\ty')\},\ek
     \beta(\tx',\ty',\ty_3)\text{ is (1,1)-periodic with respect to } \ty',&
     \text{on }\pa Z_{BL}\setminus\{\ty_3=\ga_i(\tx',\ty')\}.
  \end{array}\right.
  $$
Here $Z_{BL}$ denotes the semi-infinite cylinder $\{(\ty',\ty_3)\in\R^3:\ty'\in (0,1)\ti
(0,1), \ty_3<\ga_i(\tx',\ty')\}$,
and $[\varphi(x)]_S:=\lim_{s\downarrow
0}(\vp(x+s e_3)-\vp(x-s e_3))$ is the jump at
$S=(0,1)\ti(0,1)\ti\{0\}$. In this subsection the unit vector in the
direction of $\ty_l$-axis is denoted by $e_l$ for $l=1\sim 3$.

We assume w.l.o.g. that $D\varphi_i(\tx')\in C^1(\bar{U}_i)$,
$(D\varphi_i(\tx'))^{-1}\in C^1(\bar{V}_i)$.
\par
Define the space $\V$ by
 $$\begin{array}{l}
\V:=\{v\in L^2_{\text{loc}}(Z_{BL}):\na v\in L^2(Z_{BL}),
\div_\ty(B_i(\tx')^T v)=0,\ek
 \hspace{1cm} v(\cdot,\ga_i(\tx',\cdot))=0 \text{ (in a trace sense)},
 v\text{ is (1,1)-periodic w.r.t. $\ty'$}\}
  \end{array}$$
 endowed with norm $\|v\|_{\V}:=\|\na v\|_{L^2(Z_{BL})}$. Then $\V$  is a Banach space.
  \par
  Testing $\PP$ formally with $\varphi\in\V$, one gets the equality
\begin{equation}\label{E3.7}
(B_i(\tx')\na_\ty\beta(\tx',\ty),B_i(\tx')\na_\ty
\varphi)_{Z_{BL}}=-\int_S\la\cdot\varphi\, d\ty'.
\end{equation}

  \begin{tdefi}\label{D3.3}
  \rm{
 A function $\beta=\beta_i(\tx',\ty,\la)\in\V$ is called a solution
 to $\PP$ if it satisfies \eq{3.7} for all $\varphi\in\V$.
  }
  \end{tdefi}

  \begin{theo}
  \label{T3.4}
  {\rm
  There exists a unique solution to the problem $\PP$ in the sense of Definition \ref{D3.3}.}
  \end{theo}
\textbf{Proof.}
 Note that, by Poincar\'e's inequality,
$$|\int_S\la\cdot\varphi\, d\ty'|\leq
|\la|\|\varphi\|_{L^2(S)}\leq c|\la|\|\na\varphi\|_{L^2(Z^{+}_{BL})} \leq c|\la|\|\varphi\|_{\V},$$
where and in what follows $Z^{+}_{BL}=\{\ty\in Z_{BL}:\ty_3>0\}$. Thus, by Lax-Milgram's lemma we get the conclusion.\hfill $\Box$
\par
We give a variation of
De-Rham's lemma without proof, that can be easily proved using standard techniques.
\begin{lem}
\label{L3.5} {\rm Let a matrix $B$ be nonsingular and suppose that
$h\in H^{-1}(Z_{BL})$ satisfies
$$\lan h,v\ran_{H^{-1},H^1_0}=0$$
  for all
$v\in H^1_0(Z_{BL})$ with $\div(B^Tv)=0$. Then
\begin{equation}
\label{E3.8n} h= B\na\varphi
\end{equation}
with some unique $\varphi\in L^2_{(m)}(Z_{BL})$. }
\end{lem}

\begin{rem}
\label{R3.6} {\rm  If $\beta$ is a solution to $\PP$ in the sense of
Definition \ref{D3.3}, then it follows by Lemma \ref{L3.5} and
integration by parts that there is some
$\om=\om_i(\tx',\cdot,\la)\in L^2_{(m)}(Z_{BL})$ such that
$\{\beta,\om\}$ solves the first equation of $\PP$. On the other
hand, one can easily verify that the first and second equations of
$\PP$ form an elliptic system in the sense of Agmon, Douglis and
Nirenberg. Then, by the interior regularity for solutions to ADN
elliptic systems (cf. \cite{ADN64},  Theorem 10.3), we get
 \begin{equation}
 \label{E3.11}
\{\beta,\om\}\in \big(\V\cap C^\infty(Z_{BL}\setminus S)^3\big)\ti
\big(L^2_{(m)}(Z_{BL})\cap C^\infty(Z_{BL}\setminus S)\big).
 \end{equation}
 Moreover, it follows that $\{\beta,\om\}$ satisfies the fourth equation of $\PP$,
 that is, the jump condition. In fact, since
 $$\div_\ty\big(-A_i(\tx')\na_\ty\beta(\tx',\ty)+B_i(\tx')^T\om(\tx',\ty)\big)=0,\quad \text{in
 }Z_{BL}\setminus S,
$$
and
$$-A_i(\tx')\na_\ty\beta(\tx',\ty)+B_i(\tx')^T\om(\tx',\ty)\in
L^2(Z_{BL}),$$ one gets that for any Lipschitz subdomain $G$ of
$Z_{BL}\setminus S$
 $$\big(-A_i(\tx')\na_\ty\beta(\tx',\ty',\ty_3)+B_i(\tx')^T\om(\tx',\ty',\ty_3)\big)\cdot
 {\bf n}
 \in H^{-1/2}_{loc}(\pa G)$$ where ${\bf n}$ is the outward normal vector at
 the boundary $\pa G$, see \cite{Te77} or \cite{Ga94-1}.
 Now,
testing the first equation of $\PP$ with $\varphi\in\V\cap
C^\infty_0(Z_{BL})$ yields
$$
\lan\big[\frac{\pa\beta}{\pa\ty_3}-\om\nu(\vp_i(\tx'))\big]_S,\varphi\ran_{H^{-1/2}(S),
H^{1/2}(S)} =\int_S\la\cdot\varphi\,d\ty'$$ implying the fourth
equation of $\PP$.

Furthermore, testing the first equation of $\PP$ with suitable
functions, one can easily see that $\om$ is (1,1)-periodic with
respect to $\ty'$ provided its trace has a meaning.

Henceforth we shall call $\{\beta,\om\}$ a solution to $\PP$ as
well.}
\end{rem}
\begin{rem}
\label{R3.7}
\rm{
Given $b<0$, let $\{\beta^{(b)},\om^{(b)}\}$ be the unique solution
to the problem $\PP$ with jump conditions at $S$ replaced by the ones at $S(b):=\{\ty_3=b\}$.
Let us denote by $\{\beta^{(0)},\om^{(0)}\}$ the unique solution to $\PP$ with jump conditions at $S$.
Then, it is easily checked that
$$\{\beta^{(b)}(\tx',\ty), \om^{(b)}(\tx',\ty)\}=\left\{
               \begin{array}{cl}
               \{\beta^{(0)}(\tx',\ty), \om^{(0)}(\tx',\ty)\}&\quad \ty_3>0,\ek
               \{\la\ty_3+\beta^{(0)}(\tx',\ty), \om^{(0)}(\tx',\ty)\}&\quad b<\ty_3\leq 0,\ek
               \{b\la + \beta^{(0)}(\tx',\ty), \om^{(0)}(\tx',\ty)\}&\quad \ty_3\leq b.
               \end{array}
                     \right.$$

}
\end{rem}

The next theorem shows behavior of solutions to $\PP$ near the
interface $S$ and for $\ty_3\ra -\infty$.
 \begin{theo}
 \label{T3.7}
{\rm Let $\{\beta=\beta_i(\tx',\cdot,\la),
\om=\om_i(\tx',\cdot,\la)\}$ be the solution to $\PP$. Then, there
exist a constant $\alpha_i=\alpha_i(\tx')>0$ and constant vector
$c^{bl}_i=c^{bl}_i(\tx',\la)$ depending on $\ga_i$ and $\Ga$
satisfying
\begin{equation}
 \label{E3.22}
  \begin{array}{l}
  \forall \vec{k}\in \N^2_0,\quad\forall \vec{m}, \vec{l}\in\N^3_0,
  \quad \forall (\tx',\ty)\in U_i\ti \{\ty\in \Z_{BL}:\ty_3<0\};\ek
\quad |D_\la^{\vec{m}}
D^{\vec{k}}_{\tx'}D^{\vec{l}}_{\ty}\bar\beta_i(\tx',\ty',\ty_3,\la)|
+|D_\la^{\vec{m}}
D^{\vec{k}}_{\tx'}D^{\vec{l}}_{\ty}\om_i(\tx',\ty',\ty_3,\la)|\les
e^{\alpha_i \ty_3},
\end{array}
\end{equation}
and
\begin{equation}
 \label{E3.22n}
  \begin{array}{l}
  \forall \vec{k}\in \N^2_0,\,\,\forall \vec{m}\in\N^3_0,\,\,
  \forall \tx'\in U_i,\,\,\forall r\in (1,\infty);\ek
\quad
 \|D_\la^{\vec{m}} D^{\vec{k}}_{\tx'}\bar\beta_i(\tx',\ty',\ty_3,\la)\|_{W^{1,r}(Z^{+}_{BL})}
+\|D_\la^{\vec{m}}
D^{\vec{k}}_{\tx'} \om_i(\tx',\ty',\ty_3,\la)\|_{L^r(Z^{+}_{BL})} \leq C,
\end{array}
\end{equation}
where
\begin{equation}
\label{E3.10n}
\bar\beta_i(\tx',\ty,\la)\equiv\beta_i(\tx',\ty,\la)-c^{bl}_i(\tx',\la),
\end{equation}
and
$C$ depends on  the boundedness constants
of $D\vp_i, D\vp_i^{-1}$ and $\vec{l}$ and $r$ .
 }
\end{theo}
{\bf Proof.} We rely on Fourier expansion techniques. We shall write
$A=A_i,B=B_i$ for simplicity. Let $A=(a_{jl})_{j,l=1\sim 3}$.
Due to Remark \ref{R3.7}, we may assume w.l.o.g. that $\ga_i(\tx',\ty')>1$ for
all $(\tx',\ty')\in U_i\ti Z'$ where $Z'=(0,1)\ti (0,1)$.
In view of the definition of the
solution to $\PP$ and Remark \ref{R3.6}, we get that
$$\begin{array}{l}
\beta(\tx',\cdot,\la)\in C((-\infty,1],L^2(Z')),\quad
 \om(\tx',\cdot,\la)\in
L^2(-\infty,1;L^2(Z'))
 \end{array}$$
and that $\beta, \om$ as functions of $\ty'$ belong to
$C^\infty_{\text{per}}(Z')$, where $Z'=(0,1)\ti (0,1)$ and
$C^\infty_{\text{per}}(Z')$ is the subspace of $C^\infty(Z')$ formed
by all (1,1)-periodic functions. Hence we have Fourier expansions of
$\beta,\om$ such that
\begin{equation} \label{E3.11n}
 \begin{array}{l}
\beta(\tx',\ty,\la)=\di\sum_{m\in \Z^2}c_m(\tx',\ty_3,\la)
e^{2\pi\text{i}m\cdot\ty'},\quad \forall \ty_3\in (-\infty,1),\ek
 \om(\tx',\ty,\la)=\di\sum_{m\in
\Z^2}d_m(\tx',\ty_3,\la) e^{2\pi\text{i}m\cdot\ty'},\quad \text{a.a.
} \ty_3\in (-\infty,1),
\end{array}
\end{equation}
where Fourier coefficients $c_{m}=(c_{m,1},c_{m,2},c_{m,3})$, $d_{m}
(m=(m_1,m_2))$ are vector and scalar functions in $\ty_3$,
respectively. Then, using $a_{3j}=a_{j3}=0$ $(j=1,2), a_{33}=1$ we get
for $\ty_3\in (-\infty,1)\setminus \{0\}$ that
$$\begin{array}{l}
\div_\ty (A\na_\ty \beta)
=\sum_{m\in\Z^2}\Big(\frac{d^2}{d\ty_3^2}c_m-4\pi^2\xi_mc_m\Big)
e^{2\pi\text{i}m\cdot\ty'},
\end{array}
$$
where $\xi_m(\tx')\equiv \sum_{1\leq j,l\leq 2}a_{jl}(\tx')m_jm_l$.
 By positivity of the matrix $A$ there is some
 $\alpha_i=\alpha_i(\tx')>0$ satisfying
 \begin{equation}
 \label{E3.13}
  \xi_m(\tx')\geq \frac{\alpha_i^2(\tx')}{\pi^2}|m|^2,
\quad \forall m\in\Z^2.
\end{equation}
Here, without loss of generality we may regard $\alpha_i(\tx')$ as a
continuous function in $\tx'$ since $a_{jl}(\tx'), j,l=1,2,$ is
continuous in $\tx'$. Moreover, we have for $\ty_3\in
(-\infty,1)\setminus \{0\}$
 $$\begin{array}{l}
 B\na_\ty\om=\di\sum_{m\in\Z^2}
                   B\Big(
                   \begin{array}{c}
                   2\pi\text{i}d_mm\ek
                      \frac{d}{d\ty_3}d_m
                   \end{array}
                   \Big)
                   e^{2\pi\text{i}m\cdot\ty'}.
\end{array}
$$
On the other hand, $\text{div}_{\ty}(B^T\beta)=0$ implies
$\frac{d}{d\ty_3}c_m\cdot
\nu(\vp_i(\tx'))+\big((D\vp_i)^Tc_m\big)\cdot 2\pi im=0$ for all
$m\in\N_0^2$. Thus, we get the following system of ordinary
equations for each given $m\in\Z^2$:
 \begin{equation}
 \label{E3.12}
\left\{
 \begin{array}{l}
  \frac{d^2}{d\ty_3^2}c_m-4\pi^2\xi_mc_m -B\Big(
                   \begin{array}{c}
                   2\pi\text{i}d_mm\ek
                      \frac{d}{d\ty_3}d_m
                   \end{array}
                   \Big)
                   =0,\quad \text{for }\ty_3\in (-\infty,1)\setminus \{0\}\ek
\frac{d}{d\ty_3}c_m\cdot
\nu(\vp_i(\tx'))+\big((D\vp_i)^Tc_m\big)\cdot 2\pi im=0,\quad
\text{for }\ty_3\in (-\infty,1)\setminus \{0\}.
 \end{array}
 \right.
 \end{equation}
 In particular, for $m=(0,0)$ we have
\begin{equation}
\label{E3.15}
 \begin{array}{l}
  \di\frac{d^2}{d\ty_3^2}c_{(0,0)}- \frac{d}{d\ty_3}d_{(0,0)}\nu=0,
  \quad \frac{d}{d\ty_3}c_{(0,0)}\cdot\nu=0,\quad \text{for }\ty_3\in (-\infty,1)\setminus \{0\},
 \end{array}
\end{equation}
 yielding
\begin{equation} \label{E3.14}
c_{(0,0)}(\ty_3)\equiv c_i^{bl}(\tx',\la)=\text{const}, \quad
d_{(0,0)}(\ty_3)\equiv 0,\quad\forall \ty_3<0,
 \end{equation}
  in view of $\na_\ty\beta, \om\in L^2(Z'\ti(-\infty,0))$.

The solution $\{c_m, d_m\}, |m|\geq 1,$ to \eq{3.12} is found as
\begin{equation} \label{E3.25}
\begin{array}{l}
 c_m(\tx',\ty_3,\la)=
  \big(c_m^0 - 2\pi\sqrt{\xi_m}\ty_3\tilde{d}_m^0B\big(
             \begin{array}{c}
            \frac{\text{i}m}{\sqrt{\xi_m}}\\
             1
             \end{array}
              \big)
 \big) e^{2\pi\sqrt{\xi_m}\ty_3},\\[3ex]
 d_m(\tx',\ty_3,\la)=-4\pi\sqrt{\xi_m}\tilde{d}_m^0
 e^{2\pi\sqrt{\xi_m}\ty_3},\quad\forall\ty_3\in (-\infty,0),
\end{array}
\end{equation}
with
$$c_m^0=\lim_{\ty_3\ra 0}c_m(\tx',\ty_3,\la),\quad
\tilde{d}_m^0:=c_m^0\cdot B\big(
             \begin{array}{c}
            \frac{\text{i}m}{\sqrt{\xi_m}}\\
             1
             \end{array}
              \big)=c_m^0\cdot\nu(\vp_i(\tx'))
+\big((D\vp_i)^Tc_m^0\big)\cdot\frac{im}{\sqrt{\xi_m}}.$$

Now let us determine $\{\beta,\om\}$ for $\ty_3>0$. From the jump
condition $[\frac{\pa\beta}{\pa\ty_3}-\om\nu]_S=\la$ we get that
\begin{equation}
\label{E3.26}
\big[\frac{dc_m}{d\ty_3}-d_m\nu(\vp_i(\tx'))\big]_S=\left\{\begin{array}{ll}
                                               0&\quad \text{for }m\neq (0,0)\ek
                                               \la&\quad\text{for } m=(0,0).
                                                 \end{array}
                                                 \right.
\end{equation}
By \eq{3.15}, for $\ty_3\in (0,1)$ we have
\begin{equation}
\label{E3.21}
 d_{(0,0)}=
\text{const},\quad
c_{(0,0)}(\ty_3)=(d_{(0,0)}\nu(\vp_i(\tx'))+\la)\ty_3+c_i^{bl}(\tx',\la).
\end{equation}
Moreover, in view of \eq{3.26} we have \eq{3.25} for $|m|\geq 1$,
$\ty_3\in (0,1)$ as well.

Thus, we get that $c_m$, $d_m$
are defined for all $\ty_3\in (-\infty, 1)$ and
\begin{equation}
\label{E3.25n}
\begin{array}{l}
 c_m(\tx',\ty_3,\la)=
  \big(c_m^{1} - 2\pi\sqrt{\xi_m}(\ty_3-1)\tilde{d}_m^{1}B\big(
             \begin{array}{c}
            \frac{\text{i}m}{\sqrt{\xi_m}}\\
             1
             \end{array}
              \big)
 \big) e^{2\pi\sqrt{\xi_m}(\ty_3-1)},\\[3ex]
 d_m(\tx',\ty_3,\la)=-4\pi\sqrt{\xi_m}\tilde{d}_m^1
 e^{2\pi\sqrt{\xi_m}(\ty_3-1)},\quad\forall\ty_3\in (-\infty,1),
\end{array}
\end{equation}
where $c_m^1\equiv c_m(\tx',1,\la)$ and
$\tilde{d}_m^1\equiv\frac{c_m^1\cdot B \big(
             \begin{array}{c}
            \frac{\text{i}m}{\sqrt{\xi_m}}\\
             1
             \end{array}
              \big)}{1-2\pi\sqrt{\xi_m}\big[B\big(
             \begin{array}{c}
            \frac{\text{i}m}{\sqrt{\xi_m}}\\
             1
             \end{array}
              \big) \big]^2}$.

By \eq{3.25n}, for all $\ty_3<1$ we have
\begin{equation}
\label{E3.16}
 |c_m(\tx',\ty_3,\la)|+
 |\ty_3-1||d_m(\tx',\ty_3,\la)|\leq
 c|c_m^1|e^{\pi\sqrt{\xi_m}(\ty_3-1)},
\end{equation}
where the constant  $c>0$ depends on the boundedness constants of
$D\vp_i, D\vp_i^{-1}$. Moreover, $\beta$ is continuous at
$\ty_3=1$ in the norm of $L^2(Z')$ and
$$\sum_{m\in\Z^2}|c^1_m|^2=\|\beta(\tx',
\cdot,1,\la)\|^2_{L^2(Z')}.$$
  Hence, in view of \eq{3.25n}, \eq{3.16}, \eq{3.14}, we get that
$\beta-c_{(0,0)},\om-d_{(0,0)}$ are infinitely differentiable in
$U_i\ti Z_{BL} $ and
  by Parceval's equality
  that
\begin{equation}
\label{E3.18}
\begin{array}{l}
|\beta(\tx',\ty',\ty_3,\la)-c_i^{bl}(x',\la)|+|\ty_3-1||\om(\tx',\ty',\ty_3,\la)|\ek
 \hspace{1cm} =(\sum_{|m|\geq 1}|c_m(\tx',\ty_3,\la)|^2)^{1/2}
    +|\ty_3-1|(\sum_{|m|\geq 1}|d_m(\tx',\ty_3,\la)|^2)^{1/2}\ek
  \hspace{1cm} \leq c \|\beta(\tx',
\cdot,1,\la)\|^2_{L^2(Z')}e^{\al_i(\tx')(\ty_3-1)},
 \quad \forall (\tx',\ty',\ty_3)\in U_i\ti Z'\ti (-\infty,0),
\end{array}\end{equation}
and, in view of \eq{3.16}, \eq{3.21}, that
\begin{equation}
\label{E3.18n}
\begin{array}{l}
|\beta(\tx',\ty',\ty_3,\la)-c_i^{bl}(x',\la)-(d_{(0,0)}\nu(\vp_i(\tx'))+\la)\ty_3|
+|\ty_3-1||\om(\tx',\ty',\ty_3,\la)-d_{(0,0)}|\ek
  \hspace{1cm}
 \leq c \|\beta(\tx',\cdot,1,\la)\|^2_{L^2(Z')}e^{\al_i(\tx')(\ty_3-1)},
 \quad \forall(\tx',\ty',\ty_3)\in U_i\ti Z'\ti [0,1).
\end{array}\end{equation}
\par
 In the same way, using the expression \eq{3.11n}, \eq{3.25n},
we get for $\vec{l}\in\N_0^3$ with $|{\vec{l}}|\geq 1$ that
\begin{equation}
\label{E3.19}
\begin{array}{l}
 |\ty_3-1|^{\vec{l}}|D_\ty^{\vec{l}}(\beta(\tx',\ty',\ty_3,\la)-c_i^{bl}(x',\la))|
  +|\ty_3-1|^{\vec{l}+1}|D_\ty^{\vec{l}}
 \om(\tx',\ty,\la)|\leq c e^{\alpha_i(\tx')\ty_3},\ek
  \hfill\forall
 (\tx',\ty',\ty_3)\in U_i\ti Z'\ti (-\infty,0),
\end{array}
\end{equation}
and that
\begin{equation}
\label{E3.18nn}
\begin{array}{l}
|\ty_3-1|^{\vec{l}}|D_\ty^{\vec{l}}(\beta(\tx',\ty',\ty_3,\la)-c_i^{bl}(x',\la)
 -(d_{(0,0)}\nu(\vp_i(\tx'))+\la)\ty_3)|\ek
\hspace{1cm}
 +|\ty_3-1|^{\vec{l}+1}|D_\ty^{\vec{l}}(\om(\tx',\ty',\ty_3,\la)-d_{(0,0)})|\ek
  \hspace{1cm}
 \leq c \|\beta(\tx',\cdot,1,\la)\|^2_{L^2(Z')}e^{\al_i(\tx')(\ty_3-1)},
 \quad \forall (\tx',\ty',\ty_3)\in U_i\ti Z'\ti [0,1),
\end{array}\end{equation}
 with constant $c>0$ depending on the boundedness constants
of $D\vp_i, D\vp_i^{-1}$ and $\vec{l}$. In particular, \eq{3.18}
$\sim$ \eq{3.18nn} imply that
\begin{equation}
\label{E3.41}
 |D_\ty^{\vec{l}}\bar\beta(\tx',\ty',\ty_3,\la)|
   +|D_\ty^{\vec{l}}\om(\tx',\ty,\la)|\leq c e^{-\alpha_i(\tx')|\ty_3|},\quad
 \forall (\tx',\ty',\ty_3)\in U_i\ti Z'\ti (-\infty,\ha].
\end{equation}
\begin{figure}
\label{Cell}
\begin{center}
\includegraphics[scale=0.5]{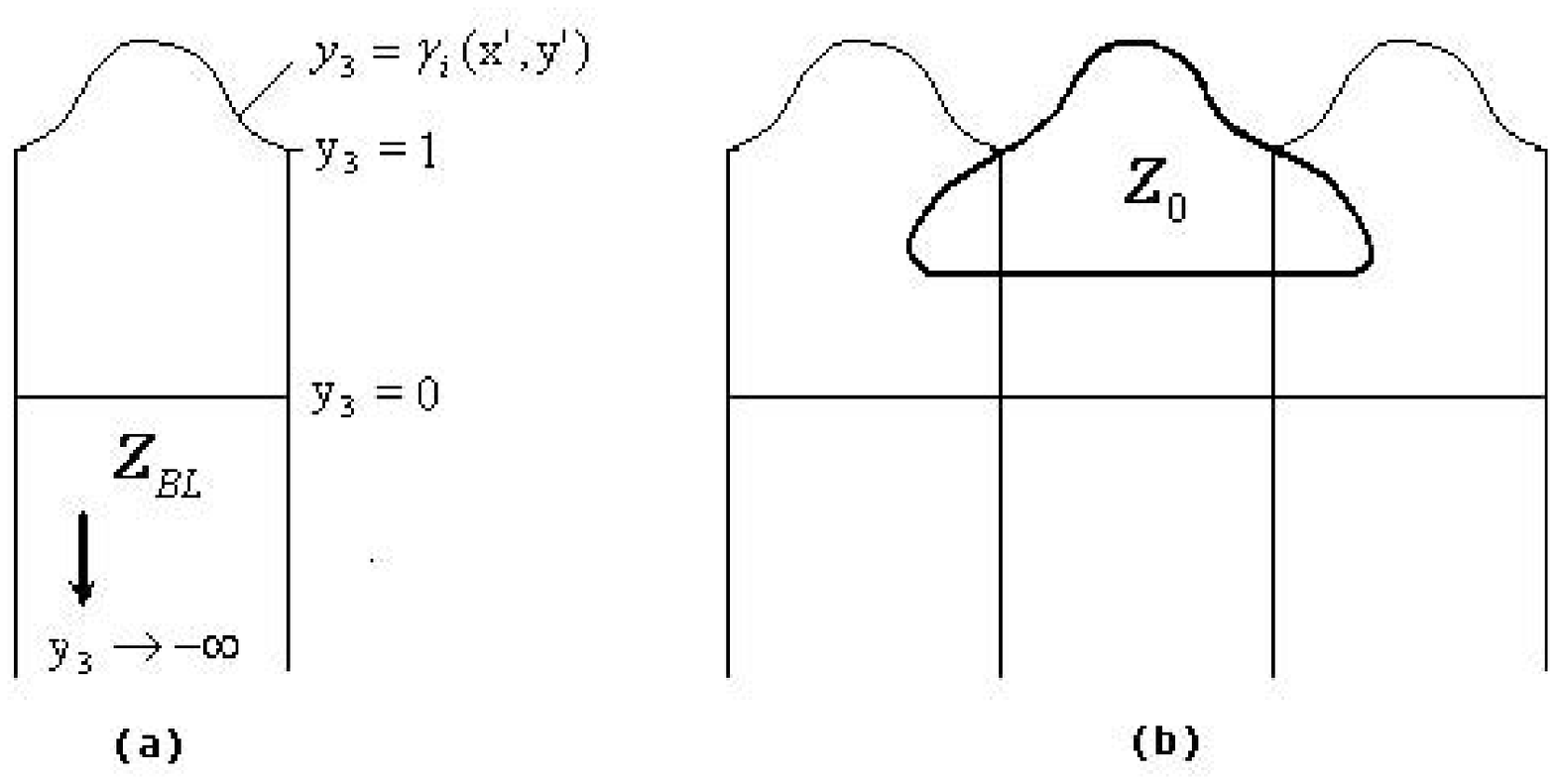}
\caption{the domains $Z_{BL}$ $\big($(a)$\big)$ and $Z_0$ $\big($(b)$\big)$}
\end{center}
\end{figure}


Thus, \eq{3.22} for $|\vec{m}|=|\vec{k}|=0$ is proved.
\par
In order to prove \eq{3.22n} for $|\vec{m}|=|\vec{k}|=0$,
 consider a smooth domain $Z_0$ expressed by Fig \ref{Cell} (b).
 By the above proved regularity, periodicity and
 continuity at $\ty_3=0$ of $\beta$, it follows that  the trace of $\beta$ on
$\pa Z_0$ belongs to $W^{1-1/r,r}(\pa Z_0)$ for any $r\in (1,\infty)$.
 Therefore, by well-known theory of existence
of solutions to inhomogeneous boundary value problems to ADN elliptic systems, see \cite{ADN64},
$$\{\beta,\om\}\in W^{1,r}(\{\ty\in Z_{BL}: \ha <\ty_3<\ga_i(\tx',\ty')\})
\ti L^{r}(\{\ty\in Z_{BL}: \ha <\ty_3<\ga_i(\tx',\ty')\}), \forall r\in (1,\infty).$$
Hence, \eq{3.22n} for $|\vec{m}|=|\vec{k}|=0$ holds true.

Now, let us show for all ${\vec{k}}\in\N_0^2$ with $|{\vec{k}}|\geq 1$ and
${\vec{l}}\in\N_0^3$ with $|{\vec{l}}|\geq 1$ that
\begin{equation}
\begin{array}{l}
\label{E3.20}
|D_{\tx'}^{\vec{k}}D_\ty^{\vec{l}}\beta(\tx',\ty,\la)|+
|D_{\tx'}^{\vec{k}}D_\ty^{\vec{l}}\om(\tx',\ty,\la)|\les
e^{\alpha_i(\tx')\ty_3},\,\, \forall (\tx',\ty',\ty_3)\in U_i\ti Z'\ti (-\infty,0).
 \end{array}
\end{equation}
Differentiating the variational equation \eq{3.7} in $\tx_j$, we get
a new variational equation with the unknown
$D_{\tx_j}\beta$ and  additional external force terms which are
exponentially decreasing. More precisely, we get
$$(B_i\na_\ty(\na_{\tx_j}\beta),B_i\na_\ty\vp)=
 -(B_i\na_\ty\beta,D_{\tx_j}B_i\na_\ty\vp)
-(D_{\tx_j}B_i\na_\ty\beta,B_i\na_\ty\vp),\quad \forall\vp\in {\cal V}.
 $$
  Then,
  $D_{\tx_j}\na_\ty\beta\in L^2(Z_{BL})$ and \eq{3.20} for
$|\vec{k}|=1$ follow in the same way as above using
Lax-Milgram's lemma and Theorem \ref{T3.4}. Then, repeating the
above argument, \eq{3.20}, \eq{3.22n} and
hence \eq{3.22}, \eq{3.22n} for for $|\vec{k}|>1$, $|\vec{m}|=0$ follows
in view of \eq{3.18}, \eq{3.19} and
$$D_\ty^{\vec{l}}\beta(\tx',\ty,\la)=D_\ty^{\vec{l}}\bar\beta(\tx',\ty,\la),|\vec{l}|\geq
1.$$

\par
 Next let us prove \eq{3.22},\eq{3.22n} for $|\vec{m}|\geq 1$.
By the end of this subsection   we use notation
\begin{equation}
\label{E3.23}
\beta^l_i(\tx',\ty):=\beta_i(\tx',\ty,e_l), \om^l_i(\tx',\ty):=\om_i(\tx',\ty,e_l),
\bar\beta^l_i(\tx',\ty):=\bar\beta_i(\tx',\ty,e_l),  l=1\sim 3.
\end{equation}
Since the mappings $\la\ra \beta_i(\tx',\ty,\la),
\la\ra \om_i(\tx',\ty,\la), \la\ra c^{bl}_i(\tx',\la)$ are linear,
we get
\begin{equation}
\label{E3.49}
\frac{\pa\bar\beta_i(\tx',\ty,\la)}{\pa\la_l}=\bar\beta^l_i(\tx',\ty),
\frac{\pa\om_i(\tx',\ty,\la)}{\pa\la_l}=\om^l_i(\tx',\ty),\,\, l=1\sim 3,
\end{equation}
 where
$\la=\la_1e_1+\la_2e_2+\la_3e_3$. Hence, by already proved
conclusion of the theorem for $|\vec{m}|=0$ we get the conclusion
for $|\vec{m}|=1$. For the case $|\vec{m}|>1$ \eq{3.22} is proved
since $\frac{\pa^2\bar\beta^\la_i}{\pa\la_h\pa\la_l}
 =\frac{\pa}{\pa\la_h}\bar\beta^l_i=0,
 \frac{\pa^2\om^\la_i}{\pa\la_h\pa\la_l}
 =\frac{\pa}{\pa\la_h}\om^l_i=0, l,h=1\sim 3$.

 The proof of the theorem is complete.
\hfill $\Box$
\begin{rem}
\label{R3.8add}
 {\rm
From Theorem \ref{T3.7} and its proof one can infer the following facts: \\
(i) If $\ga_i(x',y')\geq d\geq 0$,
$$\begin{array}{l}
  \forall \vec{k}\in \N^2_0,\,\,\forall \vec{l},
\vec{m}\in\N^3_0\,\,(|\vec{l}|\geq 1);\ek \quad\quad
|D_\la^{\vec{m}}
D^{\vec{k}}_{\tx'}D^{\vec{l}}_{\ty}\beta_i(\tx',\ty,\la)| \les
e^{\alpha_i(\tx')\ty_3},\quad (\tx',\ty)\in U_i\ti (\{\ty\in Z_{BL}:\ty_3<d/2\}\setminus S).
\end{array}
$$
In particular, for $|\vec{m}|>1$
$$D_\la^{\vec{m}}
 D^{\vec{k}}_{\tx'}D^{\vec{l}}_{\ty}\bar\beta_i(\tx',\ty,\la)=0,\quad D_\la^{\vec{m}}
 D^{\vec{k}}_{\tx'}D^{\vec{l}}_{\ty}\om_i(\tx',\ty,\la)=0,\quad (\tx',\ty)\in U_i\ti (\{\ty\in Z_{BL}:\ty_3<d/2\}\setminus S).$$
(ii) In view of
  $A_i,B_i\in
C^\infty(U_i)\cap C^1(\bar{U}_i)$, it follows that
$$c_i^{bl}(\tx',\la)=\int_S\beta_i(\tx',\ty',0,\la)\,d\ty'\in
C^\infty(U_i)\cap C^1(\bar U_i), \,\, i=1,\ldots,N.$$
 (iii) All the constants  have the order of $O(\la)$ by the
linearity of the problem $\PP$.\\
  (iv) It follows that
  $$\int_{Z'} \beta_{i}(\tx',\ty',\ty_3,\la)\cdot\nu(\vp_i(\tx'))\,d\ty'=0,
                            \quad \forall \tx'\in U_i, \forall \ty_3\leq 0, $$
  by integrating $\div(B^T\beta)=0$ in the domain
  $Z^+_{BL}$ and $Z_{BL}\setminus
  \bar{Z}^+_{BL}$
 in view of the jump condition $[\beta]_{Z'\ti\{0\}}=0$ and
 $Be_3=\nu(\vp_i(\tx'))$. In particular, for any fixed $\la\in\R^3$
\begin{equation}
\label{E3.42}
 c_i^{bl}(\tx',\la)\cdot\nu(\vp_i(\tx'))=0, \quad\tx'\in U_i, i=1,\ldots, N.
 \end{equation}
 }
\end{rem}
\par The next lemma shows additional properties of
the solution to $\PP$.
\begin{lem}
\label{L3.8} {\rm
 For $i=1,\ldots, N$ let $c^{bl}_i(\tx',e_l)=(c_{l1},c_{l2}, c_{l3})^T$ be the
constant vector for $\la=e_l, l=1\sim 3$ in Theorem \ref{T3.7}.
Then, $c_{lk}=c_{lk}$ for $l,k=1\sim 3$ and the matrix
$$\bar{C}^{bl}_i:=\left(
       \begin{array}{cc}
        c_{11} & c_{12}\\
        c_{21} & c_{22}
       \end{array}
       \right)$$
is negatively definite. }
\end{lem}
\textbf{Proof.} Let $\beta_i^l=(\beta_{i,1}^l,
\beta_{i,2}^l,\beta_{i,3}^l), l=1,2$.
We get by the definition of solution to $\PP$ that for $l,k=1\sim 3$
 $$c_{lk}=\int_{Z'}\beta^l_i\cdot e_k\,d\ty'
 =-(B_i\na_\ty \beta^k_i, B_i\na_\ty \beta^l_i)_{\Z_{BL}}$$
and, consequently,
$c_{lk}=c_{kl}$.
\par
 By the linearity of $\PP$ with respect to $\la$, for
 $\la=(\la_1,\la_2,0)^T$ one gets
 $\beta_i(\tx',\ty,\la)=\la_1\beta^1_i+\la_2\beta^2_i$, and
 $$\begin{array}{l}
\bar{C}^{bl}_i\la\la=\sum_{l,k=1}^2c_{lk}\la_l\la_k=
\int_{Z'}\sum_{l,k=1}^2\beta_{i,k}^l\la_l\la_k\,d\ty'\ek
\hspace{1.2cm} =\int_{Z'}\beta_i\cdot\la
d\ty'=-\|B_i\na_{\ty}\beta_i\|^2_{L^2(Z_{BL})}\leq 0.
 \end{array}$$
It follows by uniqueness of solution to $\PP$ that
 the equality in the above inequality holds if and only if $\la=0$.
  Therefore the matrix
$\bar{C}^{bl}_i$ is negatively definite.

Thus, the proof is complete.\hfill $\Box$
\subsection{Local boundary layer corrector}
 Using the result of boundary layer analysis, we construct a local boundary layer
 corrector in $\Ga_{\da,i}^\ve$ for $i=1,\ldots,N$. Let a three-dimensional
vector field $\la=(\la_1,\la_2,\la_3)\in C^\infty(\Ga)^3$ on $\Ga$
be given.

Define the function $\tbeta^{\ve,\la}_i: \Ga^\ve_{\da,i}\ra \R^3$ by
\begin{equation}
\label{E3.24-1}
\tbeta^{\ve,\la}_i(x)=\tbeta^{\ve,\la}_i(\Phi_i(\tx)):=
 \bar\beta_i
 (\tx',\frac{\tx}{\ve},\la\circ\varphi_i(\tx')),
 \end{equation}
where $\bar\beta_i$ is defined by \eq{3.10n} using the solution
$\beta_i(\tx',\cdot,\la\circ\varphi_i(\tx'))$ to
$(\mathbb{BL})_{\la\circ\varphi_i(\tx'),\,\tx'}$. We also define
$\tom^{\ve,\la}_i: \Ga^\ve_{\da,i}\ra \R^3$,
$\tilde{c}\,_i^{bl,\la}:\Ga^\ve_{\da,i}\ra
 \R^3$ by
\begin{eqnarray}
 \label{E3.24-2}
\tom^{\ve,\la}_i(x)=\tom^{\ve,\la}_i(\Phi_i(\tx)):=
\om_i(\tx',\frac{\tx}{\ve},\la\circ\varphi_i(\tx')),\ek
\label{E3.24-3}
\tilde{c}\,_i^{bl,\la}(x):=c_i^{bl}(\tx',\la\circ\varphi_i(\tx')).
\end{eqnarray}
 Then, $\tilde{c}\,_i^{bl,\la}\in
 C^\infty(\Ga_{\da,i}^\ve\cap\Om)^3\cap C^1(\bar\Ga_{\da,i}^\ve\cap \bar\Om)^3$,
 by Remark \ref{R3.8add} (ii) and $\tilde{c}\,_i^{bl,\la}$ is
 tangential on $\Ga$ by \eq{3.42}.

 \begin{lem}
 \label{L3.10}
 {\rm
Let $\rho(x)=d(x,\Ga)$ denote the distance from $x$ to $\Ga$.
  For all $i=1,\ldots,N$ we have:\par\noindent
 (i) $$\begin{array}{l}
  |D^{\vec{k}}_x\tbeta^{\ve,\la}_i(x)|+ |D^{\vec{k}}_x\tom^{\ve,\la}_i(x)|
 \les \ve^{-|\vec{k}|}e^{-\alpha_i(\tx') \rho(x)/\ve},
  \quad \forall x\in \Ga^\ve_{\da,i}\cap \Om, \vec{k}\in \N_0^3,\ek
  \ve\|\na\tbeta^{\ve,\la}_i(x)\|_{L^r(\Ga^\ve_{\da,i}\setminus \Om)}+
    \|\tom^{\ve,\la}_i(x)\|_{L^r(\Ga^\ve_{\da,i}\setminus \Om)}
                  \les \ve^{1/r},\quad \forall r\in (1,\infty).
  \end{array}
  $$
  (ii)
      $$\begin{array}{l}
      |-\ve\Da\tbeta^{\ve,\la}_i(x)+\na\tom^{\ve,\la}_i(x)|
       \les e^{-\alpha_i(\tx') \rho(x)/{2\ve}},\quad \forall x\in \Ga^\ve_{\da,i}\cap \Om,\ek
      \|-\ve\Da\tbeta^{\ve,\la}_i(x)+\na\tom^{\ve,\la}_i(x)\|_{L^r(\Ga^\ve_{\da,i}\setminus \Om)}
             \les \ve^{1/r},  \quad \forall r\in (1,\infty).
    \end{array}$$
  (iii)
  $$\begin{array}{l}
  |\div\tbeta^{\ve,\la}_i(x)|\les e^{-\alpha_i(\tx') \rho(x)/\ve},
  \quad \forall x\in \Ga^\ve_{\da,i}\cap \Om,\ek
  \|\div\tbeta^{\ve,\la}_i(x)\|_{L^r(\Ga^\ve_{\da,i}\setminus \Om)}\les \ve^{1/r},
                                       \quad \forall r\in (1,\infty).
    \end{array}$$
  (iv)\quad
 $[\ve\frac{\pa}{\pa\nu}\tbeta^{\ve,\la}_i(x)-\tom^{\ve,\la}_i(x)\nu(x)]_{\Ga}
   =\la(x),\quad x\in V_i$.\ek
  (v)\quad  $\tbeta^{\ve,\la}_i(x)=-\tilde{c}\,_i^{bl,\la}(x),\quad
  x\in \bar{\Ga}^\ve_{\da,i}\cap \Ga_0$.
  }
 \end{lem}
 \textbf{Proof.}
Fix any $i\in \{1,\ldots,N\}$.\par
 - {\it Proof of (i):}\par
By chain rule, for $j=1\sim 3$ we get
\begin{equation}
 \label{E3.27}
\begin{array}{l} \frac{\pa}{\pa x_j}\tbeta^{\ve,\la}_i(x)=
 \sum_{k=1}^2\frac{\pa}{\pa\xi_k}
  \bar\beta_i
   (\xi,\frac{\tx}{\ve},\la\circ\varphi_i(\tx'))|_{\xi=\tx'}\circ
\Phi_i^{-1}\cdot\frac{\pa(\Phi_i^{-1})_k}{\pa x_j}\ek
 \hspace{2cm}
+\frac{1}{\ve}\sum_{l=1}^3\frac{\pa}{\pa
\zeta_l}\bar\beta_i(\tx',\zeta,\la\circ\varphi_i(\tx'))|_{\zeta=\frac{\tx}{\ve}}
  \circ\Phi_i^{-1}\cdot\frac{\pa(\Phi_i^{-1})_l}{\pa x_j}\ek
 \hspace{2cm}
  +\sum_{m=1}^3\frac{\pa}{\pa\mu_m}
  \bar\beta_i(\tx',\frac{\tx}{\ve},\mu)|_{\mu=\la\circ\varphi_i(\tx')}
  \circ\Phi_i^{-1}\cdot\frac{\pa\la_m}{\pa x_j}.
\end{array}
\end{equation}
Also, we have the expression of the first derivatives of
$\tom^{\ve,\la}_i(x)$ in a similar form. By Theorem \ref{T3.7} for $x\in\Ga^\ve_{\da,i}\cap\Om$
both the moduli of the first and third term in the right-hand side of
\eq{3.27} are estimated by $e^{-\alpha_i(\tx')\rho(x)/\ve}$
and the modulus of the second term
 by $\ve^{-1}e^{-\alpha_i(\tx') \rho(x)/\ve}$.
 Note that $\rho(x)=\tx_3$.
Therefore the first estimate of (i) for $|\vec{k}|=1$ is proved. The first
estimate of (i) for the
cases $|\vec{k}|>1$ can be obtained by
differentiating \eq{3.27} repeatedly.\par
 The second estimate of (i) follows by \eq{3.27} and \eq{3.22n} of
 Theorem \ref{T3.7} in view of
 \begin{equation}
 \label{E3.50}
 \|h(\frac{x}{\ve})\|_{L^r(\Ga^\ve_{\da,i}\setminus \Om)}\les \ve^{1/r}, \quad \forall r\in [1,\infty),
 \end{equation}
for any $h\in L^r(\ome)$.

  - {\it Proof of (ii):}\par
Note that
 \begin{equation}
 \label{E3.31n}
\begin{array}{l}
 (D_\tx \Phi_i)^{-T}=(D_\tx
\Phi_i(\tx',0))^{-T}+R(\tx',\tx_3)=B_i(\tx')+R(\tx',\tx_3),\ek
(D_\tx \Phi_i)^{-1}(D_\tx \Phi_i)^{-T}=A_i(\tx')+S(\tx',\tx_3),
\end{array}
\end{equation}
 where matrices $R(\tx',\tx_3)$ and $S(\tx',\tx_3)$ satisfy
 $$\|R(\tx',\tx_3)\|_\infty+\|S(\tx',\tx_3)\|_\infty
 \les |\tx_3|,\, \,\,\forall \tx_3\in (-\da, \da),$$
see $\eq{3.0}$. Therefore, in view of the expression of $\Da_x$, see \eq{3.12n},
  one can get for $x\in \Ga_{\da,i}^\ve$ that
  \begin{equation}
  \label{E3.28}
 \begin{array}{l}
 \Da_x\tbeta^{\ve,\la}_i(x) = \sum_{l,k=1}^3
 A_i(\tx')_{lk}\frac{\pa^2}{\pa\tx_l\pa\tx_k}
 \bar\beta_i(\tx',\frac{\tx}{\ve},\la\circ\varphi_i(\tx'))\circ
 \Phi_i^{-1}\ek
    \hspace{1.5cm}+\sum_{l,k=1}^3 S(\tx',\tx_3)_{lk} \frac{\pa^2}{\pa\tx_l\pa\tx_k}
    \bar\beta_i
     (\tx',\frac{\tx}{\ve},\la\circ\varphi_i(\tx'))\circ
     \Phi_i^{-1}\ek
 \hspace{1.5cm}+\sum_{k=1}^3
  \frac{\pa}{\pa\tx_k}\bar\beta_i
    (\tx',\frac{\tx}{\ve},\la\circ\varphi_i(\tx'))
      \circ \Phi_i^{-1}\cdot\Da(\Phi_i^{-1})_k\ek
     \end{array}
  \end{equation}
  as in the proof of Theorem 5.1 in \cite{NNM06}.
By  Theorem \ref{T3.7} and
 $$|\tx_3|\ve^{-1}e^{-\alpha_i(\tx') |\tx_3|/\ve}\les
e^{-\alpha_i(\tx') |\tx_3|/{2\ve}}, \forall \tx_3\in \R,$$
the moduli of the second and third terms in the right-hand side of
\eq{3.28} are estimated by $O(\ve)^{-1} e^{-\alpha_i(\tx')
\rho(x)/{2\ve}}$ for $x\in \Ga_{\da,i}^\ve\cap\Om$, while
their $L^r(\Ga_{\da,i}^\ve\setminus \Om)$-norms are estimated
by $O(\ve)^{-1+1/r}$ in view of \eq{3.50}.

 Now let us expand the first term in the
right-hand side of \eq{3.28}. Direct calculation yields
  \begin{equation}
  \label{E3.29}
 \begin{array}{l}
 \sum_{l,k=1}^3
 A_i(\tx')_{lk}\frac{\pa^2}{\pa\tx_l\pa\tx_k}
 \bar\beta_i
  (\tx',\frac{\tx}{\ve},\la\circ\varphi_i(\tx'))\ek
 \hspace{1.0cm}= \frac{1}{\ve^2} \sum_{l,k=1}^3
 A_i(\tx')_{lk}\frac{\pa^2}{\pa\zeta_l\pa\zeta_k}
 \bar\beta_i
   (\tx',\zeta,\la\circ\varphi_i(\tx'))\Big|_{\zeta=\frac{\tx}{\ve}}\ek
 \hspace{1.2cm}+\frac{1}{\ve}\sum_{k=1}^2\sum_{l=1}^3
A_i(\tx')_{lk}\frac{\pa^2}{\pa\xi_k\pa\zeta_l}
 \bar\beta_i
   (\xi,\zeta,\la\circ\varphi_i(\tx'))\Big|_{\xi=\tx',\zeta=\frac{\tx}{\ve}}\ek
\hspace{1.2cm}+\sum_{l,k=1}^2
A_i(\tx')_{lk}\frac{\pa^2}{\pa\xi_k\pa\xi_l}
 \bar\beta_i
  (\xi,\frac{\tx}{\ve},\la\circ\varphi_i(\tx'))\Big|_{\xi=\tx'}\ek
\hspace{1.2cm}+\sum_{l,k=1}^2 A_i(\tx')_{lk}\frac{\pa}{\pa\tx_l}
\sum_{m=1}^3\frac{\pa}{\pa\mu_m} \bar\beta_i
(\tx',\frac{\tx}{\ve},\mu)\Big|_{\mu=\la\circ\varphi_i(\tx')}
  \cdot\frac{\pa}{\pa \tx'_k}(\la_m\circ\varphi_i(\tx')).
  \end{array}
  \end{equation}
By Theorem \ref{T3.7} and \eq{3.49}
from the second to fourth terms in the right-hand side of \eq{3.29}
are estimated by $\ve^{-1} e^{-\alpha \rho(x)/\ve}$ for $x\in \Ga_{\da,i}^\ve\cap\Om$
and have $L^r(\Ga_{\da,i}^\ve\setminus \Om)$-norm equal to $O(\ve^{1/r})$.
Thus, for $x\in \Ga_{\da,i}^{\ve}\cap\Om$ we have
\begin{equation}
\label{E3.30}
\begin{array}{l}
 \ve\Da_x\tbeta^{\ve,\la}_i(x)=\frac{1}{\ve} \sum_{l,k=1}^3
 A_i(\tx')_{lk}\frac{\pa^2}{\pa\zeta_l\pa\zeta_k}
 \bar\beta_i(\tx',\zeta)\Big|_{\zeta=\frac{\tx}{\ve},\la\circ\varphi_i(\tx')}
  \circ\Phi_i^{-1} +
 R_1\ek
 =\di\frac{1}{\ve} \div_\zeta(A_i(\tx')\na_\zeta
 \bar\beta_i(\tx',\zeta,\la\circ\varphi_i(\tx'))\Big|_{\zeta=\frac{\tx}{\ve}})
  \circ\Phi_i^{-1} +
 R_1\ek
=\frac{B_i(\tx')}{\ve}\na_\zeta\om_i
  (\tx',\zeta,\la\circ\varphi_i(\tx'))\big|_{\zeta=\frac{\tx}{\ve}}\circ\Phi_i^{-1}+R_1,
 \end{array}
\end{equation}
where $|R_1(x)|\les O(e^{-\alpha_i(\tx') \rho(x)/{2\ve}})$ for $x\in \Ga_{\da,i}^{\ve}\cap\Om$
and $\|R_1\|_{L^r(\Ga_{\da,i}^\ve\setminus \Om)}\les \ve^{1/r}$.

On the other hand, it follows from \eq{3.12n}, \eq{3.31n} that
\begin{equation}
\label{E3.31}
\begin{array}{rcl}
\na_x\tom_i^{\ve,\la}(x)
  &=&(B_i(\tx')+R(\tx',\tx_3))
 \na_\tx\om_i(\tx',\frac{\tx}{\ve},\la\circ\varphi_i(\tx'))
  \circ\Phi_i^{-1},
\end{array}
\end{equation}
where
$$\begin{array}{l}
\na_\tx\om_i
  (\tx',\frac{\tx'}{\ve},\frac{\tx_3}{\ve},\la\circ\varphi_i(\tx'))\ek
  =\left(
     \begin{array}{c}
     \na_\xi\om_i
       (\xi,\frac{\tx}{\ve},\la\circ\varphi_i)\big|_{\xi=\tx'}
     +\frac{1}{\ve}\na_{\zeta'}\om_i
      (\tx',\zeta,\la\circ\varphi_i)\big|_{\zeta=\frac{\tx}{\ve}}
      +\sum_{m=1}^3
 \om_i^{m}(\tx',\frac{\tx}{\ve})\na_{\tx'}(\la_m\circ\varphi_i)\ek
     \frac{1}{\ve}\frac{\pa}{\pa\zeta_3}
     \om_i(\xi,\zeta,\la\circ\varphi_i)\big|_{\zeta=\frac{\tx}{\ve}}
     \end{array}\right).
  \end{array}$$
Hence we get that
\begin{equation}
\label{E3.32}
\begin{array}{l}
 \na_x\tom_i^{\ve,\la}(x)\ek
    =\Big(\frac{B_i(\tx')}{\ve}\na_\zeta\om_i
  (\tx',\zeta,\la\circ\varphi_i(\tx'))\big|_{\zeta=\frac{\tx}{\ve}}
  +B'_i(\tx')\na_\xi\om_i(\xi,\frac{\tx}{\ve},\la\circ\varphi_i(\tx'))\big|_{\xi=\tx'}
       \ek
  \hspace{0.5cm}+\sum_{m=1}^3\om_i^{m}(\tx',\frac{\tx}{\ve})B'_i(\tx')\na_{\tx'}
  (\la_m\circ\vp_i(\tx')) +R(\tx',\tx_3)\na_\tx\om_i
  (\tx',\frac{\tx}{\ve},\la\circ\varphi_i(\tx'))\Big)\circ\Phi_i^{-1}.
  \end{array}
\end{equation}
By Theorem \ref{T3.7} the sum from the second to fourth term in the
bracket of the right-hand side of
 \eq{3.32} equals $O(e^{-\alpha_i(\tx') \rho(x)/{\ve}})$ for $x\in \Ga_{\da,i}^\ve\cap\Om$
 and have $L^r(\Ga_{\da,i}^\ve\setminus \Om)$-norms
 equal to $O(\ve)^{1/r}$.

Now, subtracting \eq{3.32} from \eq{3.30} yields the conclusion of
(ii).
\par
- {\it Proof of (iii):}\par
  Using the fact that divergence of a vector field is independent of the choice of
  orthogonal coordinate system, we get by Theorem \ref{T3.7} that
   $$\begin{array}{l}
   \div_x\tbeta^{\ve,\la}_i(x)=\div_{\tx}
   \bar\beta_i(\tx',\frac{\tx}{\ve},\la\circ\varphi_i(\tx'))
   \circ\Phi_i^{-1}\ek
  \hspace{0.5cm} =\big(\sum_{k=1}^2\frac{\pa}{\pa\xi_k}\bar\beta_{i,k}
   (\xi,\frac{\tx}{\ve},\la\circ\varphi_i(\tx'))|_{\xi=\tx'}
   +\frac{1}{\ve}
   \div_\zeta\bar\beta_i(\tx',\zeta,\la\circ\varphi_i(\tx'))|_{\zeta=\frac{\tx}{\ve}}\ek
   \hspace{5.2cm}  +
  \sum_{k,m=1}^3\frac{\pa}{\pa\mu_m}
    \bar\beta_{i,k}(\tx',\frac{\tx}{\ve},\mu)
       \frac{\pa\mu_m}{\pa\tx_k}|_{\mu=
       \la\circ\varphi_i(\tx')}\big)\circ\Phi_i^{-1}\ek
\hspace{0.5cm} =\big(\sum_{k=1}^2\frac{\pa}{\pa\xi_k}\bar\beta_{i,k}
   (\xi,\frac{\tx}{\ve},\la\circ\varphi_i(\tx'))|_{\xi=\tx'}
   +
  \sum_{k,m=1}^3\frac{\pa}{\pa\mu_m}
    \bar\beta_{i,k}(\tx',\frac{\tx}{\ve},\mu)
       \frac{\pa\mu_m}{\pa\tx_k}|_{\mu=
       \la\circ\varphi_i(\tx')}\big)\circ\Phi_i^{-1}
  \end{array}$$
where
$(\bar\beta_{i,1},\bar\beta_{i,2},\bar\beta_{i,3})\equiv\bar\beta_i$.
Thus, by the same argument as in the proof of (ii) we get the conclusion.

 - {\it Proof of (iv):}\par
For $x\in V_i$, we have
$$\begin{array}{l}
\Big[\ve\frac{\pa\tbeta^{\ve,\la}_i(x)}{\pa\nu}-\tom^{\ve,\la}_i(x)\nu(x)\Big]_{\Ga}\ek
\hspace{1cm}= \Big[(
\frac{\pa}{\pa\zeta_3}\beta_i(\tx',\zeta,\la\circ\varphi_i(\tx'))|_{\zeta=\frac{\tx}{\ve}}
 -\om_i(\tx',\frac{\tx}{\ve},\la\circ\varphi_i(\tx')))\nu(\vp_i(\tx'))\Big]_{S}\circ\Phi_i^{-1}\ek
\hspace{1cm}= \la\circ\varphi_i(\tx')\circ\Phi_i^{-1}=\la(x).
\end{array}
$$
\par
- {\it Proof of (v):} (v) is obvious from definition of
$\tbeta_i^{\ve,\la}$. \hfill $\Box$

 \subsection{Global boundary layer corrector}
In this subsection, a global boundary layer corrector is constructed
using cut-off functions for $\Ga$ and local boundary layer correctors
$\tbeta_i^{\ve,\la}, \tom_i^{\ve,\la}$, $i=1,\ldots, N$.
Let
 $\psi_i \in C^\infty(\bar\Ga), i=1,\ldots,N,$ be  cut-off functions such that
$$\psi_i\in C^\infty(\bar\Ga),\,\,\supp \psi_i\subset \bar{V}_i, \,\,i=1,\ldots,N,\quad
\sum_{i=1}^{N}\psi_i(x')=1,\quad x'\in \Ga,$$
 where $\{V_i\}_{i=1}^{N}$ is the open covering of $\Ga$ introduced in Section \ref{notation}.
 Let
$$\alpha:=\di\min_{i=1,\ldots,N}\min_{\tx'\in
\bar{U}_i}\alpha_i(\tx')/2.$$
Let $\tilde{\psi}_i(x)=\psi_i(x')$ for
$x\in \Ga_\da, x=\Upsilon(x',\tx_3), (x',\tx_3)\in\Ga\ti (-\da,\da)$.

 Given a three-dimensional vector field
  $\la\in C^\infty(\Ga)^3$ on $\Ga$, a global boundary layer corrector
$\{\beta^{\ve,\la},\om^{\ve,\la}\}$ on $\Ga_\da^\ve$ is defined as
\begin{equation}
\label{E3.33}
\beta^{\ve,\la}(x):=\di\sum_{i=1}^N\tilde\psi_i(x)\tbeta^{\ve,\la}_i(x),\quad
\om^{\ve,\la}(x):=\di\sum_{i=1}^N\tilde\psi_i(x)\tom^{\ve,\la}_i(x),\quad
x\in\Ga_\da^\ve,
\end{equation}
 where $\tbeta^{\ve,\la}_i,\tom^{\ve,\la}_i$ are given by \eq{3.24-1}, \eq{3.24-2},
 respectively.
 Furthermore, a vector function $c^{bl,\la}: \Ga_\da^\ve\ra\R^3$ is given by
$$c^{bl,\la}(x):=\di\sum_{i=1}^N\psi_i(x)\tilde{c}\,^{bl,\la}_i(x),\,\,
x\in \Ga_\da^\ve,$$
with $\tilde{c}\,^{bl,\la}_i$ given by \eq{3.24-3}.
 Then,
 \begin{equation}
\label{E3.34}
  c^{bl,\la}\in C^\infty(\Ga_\da^\ve)^3\cap C^1(\bar\Ga_\da^\ve\cap\bar\Om)^3,\quad
   c^{bl,\la}|_\Ga\in C^\infty(\Ga)^3\cap C^1(\bar\Ga)^3,
   \end{equation}
and $c^{bl,\la}$ is tangential on $\Ga$.

\begin{lem}
\label{L3.9} {\rm
(i) It holds
$$|D^{\vec{k}}\beta^{\ve,\la}(x)|+|D^{\vec{k}}\om^{\ve,\la}(x)|\les
   \ve^{-|\vec{k}|}e^{-\alpha \rho(x)/\ve},\quad
   \forall x\in \Ga_\da^\ve\cap \Om, \forall \vec{k}\in \N_0^3,$$
and
$$\ve\|\na\beta^{\ve,\la}(x)\|_{L^r(\Ga_\da^\ve\setminus\Om)}
 +\|\om^{\ve,\la}(x)\|_{L^r(\Ga_\da^\ve\setminus\Om)}\les
   \ve^{1/r},\quad  \forall r\in (1,\infty).$$
(ii) It holds
$$|-\ve\Da\beta^{\ve,\la}(x)+\na\om^{\ve,\la}(x)|\les
e^{-\alpha \rho(x)/\ve},\quad \forall x\in \Ga_\da^\ve\cap \Om,$$
and
$$\|-\ve\Da\beta^{\ve,\la}(x)+\na\om^{\ve,\la}(x)\|_{L^r(\Ga_\da^\ve\setminus\Om)}\les
   \ve^{1/r},\quad  \forall r\in (1,\infty).$$
(iii) It holds
$$|\div\beta^{\ve,\la}(x)|\les e^{-\alpha
\rho(x)/\ve},\quad \forall x\in \Ga_\da^\ve\setminus \Ga,$$
and
$$\|\div\beta^{\ve,\la}(x)\|_{L^r(\Ga_\da^\ve\setminus\Om)}\les
   \ve^{1/r},\quad  \forall r\in (1,\infty).$$
(iv)\quad $\big[
 \ve\frac{\pa}{\pa\nu}\beta^{\ve,\la}(x)
  -\om^{\ve,\la}(x)\nu(x)\big]_{\Ga}=\la$.
}
\end{lem}
\textbf{Proof.} - {\it Proof of (i):}
\par
Note that
$$D^{\vec{k}}\beta^{\ve,\la}(x)=\di\sum_{i=1}^N
 \sum_{\vec{k_1}+\vec{k_2}=\vec{k}}
 D^{\vec{k_1}}\tilde{\psi}_i(x)D^{\vec{k_2}}\tbeta^{\ve,\la}_i(x),
$$
$$D^{\vec{k}}\om^{\ve,\la}(x)=\di\sum_{i=1}^N
 \sum_{\vec{k_1}+\vec{k_2}=\vec{k}}
 D^{\vec{k_1}}\tilde{\psi}_i(x)D^{\vec{k_2}}\tom^{\ve,\la}_i(x).
$$
Hence, by Lemma \ref{L3.10} we get the conclusion (i).
\par
- {\it Proof of (ii):}
\par
 Direct calculations yield that
$$\begin{array}{l}
-\ve\Da\beta^{\ve,\la}(x)+\na\om^{\ve,\la}(x)= \di\sum_{i=1}^N
\tilde{\psi}_i(x)(-\ve\Da\tbeta^{\ve,\la}_i(x)+\na\tom^{\ve,\la}_i(x))\ek
\hspace{3cm}+\text{ derivatives of $\tbeta^{\ve,\la}_i$ up to first
order multiplied by }\ve\ek
 \hspace{3cm} +\text{ zeroth derivative terms of } \tom^{\ve,\la}_i.
\end{array}$$
Thus, from Lemma \ref{L3.10} (ii) we get the conclusion (ii).
\par
- {\it Proof of (iii):} \par The conclusion (iii) follows directly from
Lemma \ref{L3.10} (iii) since
$$\div\beta^{\ve,\la}(x)=
\di\sum_{i=1}^N( \tilde{\psi}_i(x)\div\tbeta^{\ve,\la}_i(x)
+\text{zeroth derivative terms of $\tbeta^{\ve,\la}_i$}).$$
\par
 - {\it Proof of (iv):} \par
The conclusion (iv) is obvious
since $\tilde{\psi}_i(x),x\in\Ga_\da,$ depends only on tangential
variables of $\Ga$ and hence
 $\frac{\pa}{\pa\nu}\tilde{\psi}_i(x)=0$ .
  \hfill $\Box$
\subsection{Construction of first order approximations and Navier wall-laws}
The global boundary layer corrector constructed above rapidly
decreases with exponential decay rate going from $\Ga$ to the
interior of $\Om$. Using the corrector we construct higher
order approximations for the real solution $u^\ve$. Then, we
derive an effective Navier wall-law.
\par
 Let us fix a vector field  $\la^{(l)}\in C^\infty(\Ga)^3, l=1\sim 3,$
on  $\Ga$ with
 $$|\la^{(l)}(x')|=1, \la^{(l)}(x')\bot\la^{(k)}(x') \,\, (l\neq k),\quad
x'\in\Ga,$$ and $\la^{(3)}(x')=\nu(x')$.
\par
For $x\in \Ga_\da^\ve$, $x=\Upsilon(x',\tx_3)$, let $\la(x)\equiv
\la(x')$,
 \begin{equation}
 \label{E3.35n}
\beta^{\ve,\,l}(x)\equiv \beta^{\ve,\,\la^{(l)}}(x),
\om^{\ve,\,l}(x)\equiv \om^{\ve,\,\la^{(l)}}(x),\quad l=1\sim 3,
 \end{equation}
see \eq{3.33}, and let
\begin{equation}
\label{E3.56}
c_{lk}(x'):=c^{bl,\la^{(l)}}(x')\cdot\la^{(k)}(x'),x'\in\Ga,\quad l,k=1\sim 3,
\end{equation}
  see \eq{3.34}.
Note that
$$c_{l3}(x)=0,\quad l=1\sim 3,$$ since
$c^{bl,\la^{(l)}}(x'),l=1\sim 3,$ is tangential on $\Ga$.

Now, define $2\ti 2$ matrix $c^{bl}(x')$ by
\begin{equation}
\label{E3.46}
 c^{bl}(x')=\left(
         \begin{array}{cc}
         c_{11}(x') & c_{12}(x')\ek
         c_{21}(x') & c_{22}(x')\ek
         \end{array}
         \right),\quad x'\in\Ga.
\end{equation}
 Then, by Lemma \ref{L3.8} the matrix $c^{bl}(x')$ for all $x'\in
\Ga$ is negatively definite and $c^{bl}\in C^1(\bar\Ga)$.
 The extension of $c^{bl}$
 by zero matrix on $\Ga_1$ is denoted again by $c^{bl}$.

 Let us take a function $\Psi\in C^1(\bar\Ga)$ and its extension
$\tilde{\Psi}\in W^{1,\infty}(\Ga_\da^\ve)$ such that
 $$\begin{array}{l}
\hspace{-2cm} \Psi(x')\equiv 1\,\,\,\text{for }x'\in\Ga,\quad
\tilde{\Psi}(x)\equiv 1\,\,\,\text{for }x\in\Ga_\da^\ve
 \end{array}$$
 if $\Ga$ and
$\Ga_1$ are components of $\pa\Om$. If $\Ga$ and
$\Ga_1$ are adjacent, then we take a function $\Psi\in C^1(\Ga)$
satisfying
 $$\left\{\begin{array}{l}
 0\leq \Psi(x')\leq 1\,\,x'\in \Ga, \quad \Psi(x')\equiv 1
 \,\,\,x'\in \Ga',\quad\supp\Psi=\bar\Ga,\ek
\Psi(x')\sim \ve^{-1}d(x',\bar\Ga\cap\Ga_1),\quad
|D_{x'}\Psi(x')|\les \ve^{-1}\quad\text{for } x'\in
\Ga\setminus\Ga'.
\end{array}\right.
 $$
 In order to take a suitable extension $\tilde\Psi$ of $\Psi$ onto $\Ga_\da^\ve$, let us
 choose a domain $D\subset \Ga_\da^\ve$ such that
$D\subset\Ga_\da^\ve\cap\Om$ and
$$\pa D\cap \Ga=\Ga',\quad \text{dist}(x,\pa\Ga_\da^\ve\setminus\Ga_0)\gtrsim \ve+k\tx_3^{1/4}
\quad\text{for } x=\Upsilon(x',\tx_3)\in \pa D\cap \Om$$
 with some constant $k>0$. Then we choose a function $\tilde\Psi$ such that
$$\tilde\Psi(x)=0\quad\text{for } x\in (\Ga_\da^\ve\cap\Om)\setminus D,
\quad |\na\tilde\Psi(x)|\leq \frac{K}{\ve+\tx_3^{1/4}}\quad\text{for
} x\in \Ga_\da^\ve\cap\Om,$$ and $\tilde\Psi(x)\equiv\Psi(x')$ for
$x=\Upsilon(x',\tx_3)\in\ome\setminus\Om$. Here the constant $K$
depends on $\da,\Ga$. Obviously,
$$\|\na\tilde\Psi\|_{L^\infty(\ome\setminus\Om)}\les
\frac{1}{\ve},\quad \na\tilde\Psi\equiv 0 \quad \text{in
}(\ome\setminus\Om)\cap\Ga'_\da.
$$
\par
\begin{lem}
\label{L3.11n} {\rm Let $q>2$ if $\Ga_0$ and $\Ga_1$ are components
of $\pa\ome$ (equivalently, $\Ga$ and $\Ga_1$ are components of
$\pa\Om$) and $q>3$ if $\Ga_0$ and $\Ga_1$ are adjacent
(equivalently, $\Ga$ and $\Ga_1$ are adjacent). Then, for all $v\in
W^{1,q}(\Ga_\da^\ve)$ the following inequality holds:
\begin{equation}
\label{E3.34nn}
\begin{array}{l}
 \|\na(\tilde\Psi v)\|_{L^{2}(\Ga_\da^\ve)}\les
\|v\|_{W^{1,q}(\Ga_\da^\ve)}.
 \end{array}
\end{equation}
}
\end{lem}
 {\bf Proof.}  The proof for the case where $\Ga_0$ and $\Ga_1$ are components of $\pa\Om$ is trivial.

  Let $\Ga_0$ and $\Ga_1$ be adjacent. Then, in view of the construction of
 $\tilde{\Psi}$ we get that
\begin{equation}
\label{E3.34n}
\begin{array}{l}
 \|\na(\tilde\Psi v)\|_{L^{2}(\Ga_\da^\ve)}\leq
 \|\tilde\Psi\na v\|_{L^{2}(\Ga_\da^\ve)}+\|\na\tilde\Psi v\|_{L^{2}(\Ga_\da^\ve\cap\Om)}
 +\|\na\tilde\Psi v\|_{L^{2}(\ome\setminus\Om)}\ek
  \les \|v\|_{W^{1,2}(\Ga_\da^\ve)}+\|\tx_3^{-\frac{1}{4}}v\|_{L^{2}(\Ga_\da^\ve\cap\Om)}
 +\|\na\tilde\Psi (v-\bar{v})\|_{L^{2}(\ome\setminus\Om)}
 + |\bar{v}|\|\na\tilde\Psi\|_{L^{2}(\om)},
 \end{array}
\end{equation}
where
$\bar{v}=\frac{1}{|\ome\setminus\Om|}\int_{\ome\setminus\Om}v\,dx$
and $\om:=\ome\setminus(\Om\cup\Ga'_\da)$. Note that $|\om|=O(\ve^2)$.
The second term in the right-hand side of \eq{3.34n} is estimated as
$$\|\tx_3^{-\frac{1}{4}}v\|_{L^{2}(\Ga_\da^\ve\cap\Om)}\leq
\|\tx_3^{-\frac{1}{4}}\|_{L^3(\Ga_\da^\ve)}\|v\|_{_{L^6(\Ga_\da^\ve)}}
\les \|v\|_{W^{1,2}(\Ga_\da^\ve)}$$ using H\"older's inequality and
Sobolev embedding theorem, and the third term as
$$\|\na\tilde\Psi
(v-\bar{v})\|_{L^{2}(\ome\setminus\Om)}\leq\frac{1}{\ve}
\|v-\bar{v}\|_{L^2(\ome\setminus\Om)}\les \|\na
v\|_{L^{2}(\ome\setminus\Om)}$$ using Poincar\'e's inequality.
Finally, the fourth term in the right-hand side of \eq{3.34n} is
estimated as
$$|\bar{v}|\|\na\tilde\Psi\|_{L^{2}(\om)}\les
\|v\|_{L^\infty(\ome\setminus\Om)}\cdot \frac{1}{\ve}|\om|^{1/2}\les
\|v\|_{W^{1,q}(\Ga_\da^\ve)}$$ with the help of Sobolev embedding
$W^{1,q}(\Ga_\da^\ve)\hookrightarrow L^\infty (\Ga_\da^\ve)$ due to
$q>3$.

Thus \eq{3.34nn} is proved. \hfill $\Box$
\par
Now, let $\frac{\widetilde{\pa u}}{\pa\nu}, \tilde{\tilde{p}}\in
W^{1,q}(\Om^\ve)$ be respectively some extensions of $\frac{\pa
u}{\pa \nu}|_\Ga, p|_\Ga$ given by a linear bounded extension operator from $W^{1-1/q,q}(\Ga)$ to
$W^{1,q}(\Ga_\da^\ve)$ such that
\begin{equation}
\label{E3.36}
\begin{array}{l}
  \ttp=0,\frac{\widetilde{\pa u}}{\pa\nu}=0,\quad x\in \Om^\ve\setminus \Ga^\ve_\da,\ek
 \|\ttp,\frac{\widetilde{\pa u}}{\pa\nu}\|_{W^{1,q}(\Ga^\ve_\da)}
\les \|u\|_{W^{2,q}(\Om)}+ \|p\|_{W^{1,q}(\Om)}.
\end{array}
\end{equation}
The existence of such extension operator can
be shown by Sobolev extension theorem using the assumption on $\Ga$.
In the sequel, we use the notation
\begin{equation}
\label{E3.51}
\chi_l(x):=\Psi(x)\frac{\widetilde{\pa u_\tau}}{\pa\nu}(x)\cdot\la^{(l)}(x'),l=1,2,
\quad \chi_3(x):= -\Psi(x)\ttp(x),\quad x\in\Ga_\da^\ve,
\end{equation}
 where
$\frac{\pa u_\tau}{\pa\nu}$ denotes the $\nu$-directional derivative
of
$$u_\tau(x)=u_\tau(\Upsilon(x',\tx_3)):=u(x)-u_\nu(x)\nu(x'), \quad
u_\nu(x):=u(x)\cdot\nu(x').$$
 Note that $\frac{\pa u_\tau(x)}{\pa\nu}$ on $\Ga'$ is tangential on $\Ga'$ since
  $(\frac{\pa u}{\pa\nu})\cdot\nu= \frac{\pa u_\nu}{\pa\nu}-u_\nu=0$
  in view of the solenoidal condition for $u$ and $u|_{\Ga'}=0$.
We put $\chi:=\sum_{l=1}^3\chi_l\la^{(l)}$.

  We construct a correction $\eta^\ve$ rapidly oscillating in a neighborhood of
$\Ga$ by
\begin{equation}
\label{E3.38}
 \eta^\ve(x):=\ve\sum_{l=1}^3 \beta^{\ve,l}(x)\chi_l(x).
x\in\Ga_\da^\ve,
\end{equation}
Note that the function $\eta^\ve$ after extended by $0$ to $\ome$
belongs to $W^{1,q}(\ome)$. Moreover, if $q\geq 2$ is
given as in Lemma \ref{L3.11n}, then by \eq{3.36} and Sobolev
embedding theorem one has
\begin{equation}
\label{E3.44n}
 \|\chi_l\|_{W^{1,q}(\Ga_\da^\ve)}\les
\|u\|_{W^{2,q}(\Om)}+ \|p\|_{W^{1,q}(\Om)},\quad l=1\sim 3.
  \end{equation}
\par In order to construct a non-oscillating correction,
 consider the following problem:
\begin{equation}
\label{E3.35}
\begin{array}{rccl}
 -\Da\eta+\na\zeta &= & 0 &  \text{in }\Om,\ek
 \div \eta & = &0 &\text{in }\Om,\ek
 \eta_\tau &= & \Psi c^{bl}\frac{\pa u_\tau}{\pa\nu} &
   \text{on }\Ga,\ek
 \eta_\nu & = & 0 & \text{on }\Ga,\ek
 \eta & = & 0 & \text{on }\Ga_1,
\end{array}
\end{equation}
where and in what follows
 \begin{equation}
 \label{E3.57}
  c^{bl}\frac{\pa u_\tau}{\pa\nu}\equiv
  \sum_{l,k=1}^2c_{lk}(\frac{\pa u_\tau}{\pa\nu})_k\la^{(l)},
    (\frac{\pa u_\tau}{\pa\nu})_k\equiv
\frac{\pa u_\tau}{\pa\nu}\cdot \la^{(k)}.
\end{equation}
\par  The system \eq{3.35} has a unique
weak solution $\{\eta,\zeta\}\in W^{1,2}(\Om)\ti L^{2}_{(m)}(\Om)$
since the boundary data for
$\eta$ belongs to $H^{1/2}(\pa\Om)$, and, in view of \eq{3.34}, \eq{3.44n},
we have
 \begin{equation}
  \label{E3.35nn}
\begin{array}{rcl}
 \|\eta\|_{W^{1,2}(\Om)}+\|\zeta\|_{L^{2}_{(m)}(\Om)}&\les&
\|\Psi(\cdot)c^{bl}(\cdot)\frac{\pa
u_\tau}{\pa\nu}\|_{H^{1/2}(\Ga)}\ek
 &\leq&
c(\Om)\|\tilde\Psi(\cdot)c^{bl}(\cdot)\widetilde{\frac{\pa
u_\tau}{\pa\nu}}\|_{W^{1,2}(\Om\cap\Ga_\da^\ve)}\ek
 &\les& \|u\|_{W^{2,q}(\Om)}
 \end{array}
 \end{equation}
provided $q$ is given as in Lemma \ref{L3.11n}.

\par Let us construct
a correction $\bar\eta^\ve$ non-oscillating in a neighborhood of
$\Ga$ by
\begin{equation}
\label{E3.37}
 \bar\eta^\ve(x)=\left\{
    \begin{array}{cl}
    \ve\eta(x), & \text{for }x\in \Om,\ek
    \ve \tilde\Psi(x)c^{bl}(x)\frac{\widetilde{\pa u_\tau}}{\pa\nu}, & \text{for }x\in
\Om^\ve\setminus \Om.
\end{array}\right.
\end{equation}
Note that $\div(\bar\eta^\ve+\eta^\ve)\neq 0$, in general, and
 $(\bar\eta^\ve+\eta^\ve)|_{\pa\Om^\ve}=0$ in view of  Lemma \ref{L3.10} (v) and Remark \ref{R3.8add} (iv).

\begin{lem}
\label{L3.15}
{\rm
The vector $c^{bl}\frac{\pa u_\tau}{\pa\nu}$ on $\Ga$ defined by \eq{3.57} and
the approximations $\bar\eta^\ve$ and $\eta^\ve$ defined by \eq{3.37}, \eq{3.38}, respectively,
 are independent of the choice of  orthogonal tangent vector fields on $\Ga$.
}
\end{lem}
{\bf Proof:}  Let $\{\la^{(1)},\la^{(2)}\}$ and $\{\xi^{(1)},\xi^{(2)}\}$
 be different curvilinear systems of orthogonal tangential vector fields
on $\Ga$.
  For $x'\in\Ga$ we denote the rotational matrix from $\{\xi^{(1)}(x'),\xi^{(2)}(x')\}$
  to $\{\la^{(1)}(x'),\la^{(2)}(x')\}$ by
         $N=\Big(\begin{array}{cc}
                    n_1 & n_2\\
                    -n_2 & n_1
                  \end{array}
             \Big)$, where $n_1^2+n_2^2=1$,
 i.e.,
 $$\Lambda=N \Xi, \quad\Lambda:=(\la^{(1)}(x'), \la^{(2)}(x'))^T, \Xi:=(\xi^{(1)}(x'), \xi^{(2)}(x'))^T.$$
 Let $C=(c_{ij})$ and $D=(d_{ij})$ denote
 the $2\ti 2$ matrices defined by \eq{3.46}
  corresponding to $\{\la^{(1)},\la^{(2)}\}$ and $\{\xi^{(1)},\xi^{(2)}\}$, respectively.
When $A_j,B_j$, $j=1,2$, are two dimensional vectors, we use short notation
 $$\Big(\begin{array}{c}
         A_1\\
         A_2
         \end{array}
 \Big): \Big(\begin{array}{c}
         B_1\\
         B_2
         \end{array}
 \Big):= \Big(\begin{array}{c}
         A_1\cdot B_1+A_1\cdot B_2\\
         A_2\cdot B_1+A_2\cdot B_2
         \end{array}
 \Big).$$
 Then, it is easily seen that
 $$\Big[N\Big(\begin{array}{c}
         A_1\\
         A_2
         \end{array}
 \Big)\Big]: \left(\begin{array}{c}
         B_1\\
         B_2
         \end{array}
 \right)=
 N\Big[\Big(\begin{array}{c}
         A_1\\
         A_2
         \end{array}
 \Big): \Big(\begin{array}{c}
         B_1\\
         B_2
         \end{array}
 \Big)\Big].$$
When $x'=\vp_i(\tx')$, let
 $$C^{bl}(\Lambda)(x'):=\Big(\begin{array}{c}
         c^{bl}(\tx', \la^{(1)}\circ \vp_i(\tx'))\\
         c^{bl}(\tx', \la^{(2)}\circ \vp_i(\tx'))
         \end{array}
 \Big),\quad
  C^{bl}(\Xi)(x'):=\Big(\begin{array}{c}
         c^{bl}(\tx', \xi^{(1)}\circ \vp_i(\tx'))\\
         c^{bl}(\tx', \xi^{(2)}\circ \vp_i(\tx'))
         \end{array}
 \Big).$$
Note that $C^{bl}(\Lambda)=N C^{bl}(\Xi)$ holds by the linearity of $c^{bl}(\tx',\la)$ w.r.t. $\la$.
Hence, in view of $N^TN=I$, we have
$$\begin{array}{rcl}
\sum_{l,k=1}^2c_{lk}(\frac{\pa u_\tau}{\pa\nu}\cdot \la^{(k)})\la^{(l)}
 &=& \Lambda^T \Big[ C^{bl}(\Lambda):
       \Big(\begin{array}{c}
         \la^{(1)}(\frac{\pa u_\tau}{\pa\nu}\cdot \la^{(1)})\ek
         \la^{(2)}(\frac{\pa u_\tau}{\pa\nu}\cdot \la^{(2)})
         \end{array}
      \Big)\Big]\ek
      &=&\Xi^T N^T N\Big[ C^{bl}(\Xi):
       \Big(\begin{array}{c}
         \la^{(1)}(\frac{\pa u_\tau}{\pa\nu}\cdot \la^{(1)})\ek
         \la^{(2)}(\frac{\pa u_\tau}{\pa\nu}\cdot \la^{(2)})
         \end{array}
      \Big)\Big]\ek
      &=&\Xi^T \Big[ C^{bl}(\Xi):
       \Big(\begin{array}{c}
         \la^{(1)}(\frac{\pa u_\tau}{\pa\nu}\cdot \la^{(1)})\ek
         \la^{(2)}(\frac{\pa u_\tau}{\pa\nu}\cdot \la^{(2)})
         \end{array}
      \Big)\Big].
  \end{array}$$
Here, the first component of the vector $\Big[ C^{bl}(\Xi):
       \left(\begin{array}{c}
         \la^{(1)}(\frac{\pa u_\tau}{\pa\nu}\cdot \la^{(1)})\\
         \la^{(2)}(\frac{\pa u_\tau}{\pa\nu}\cdot \la^{(2)})
         \end{array}
      \right)\Big]$ is calculated as
$$\begin{array}{l}
\Big[ C^{bl}(\Xi):
       \Big(\begin{array}{c}
         \la^{(1)}(\frac{\pa u_\tau}{\pa\nu}\cdot \la^{(1)})\\
         \la^{(2)}(\frac{\pa u_\tau}{\pa\nu}\cdot \la^{(2)})
         \end{array}
      \Big)\Big]_1
\ek
\hspace{2cm}=c^{bl}(\tx', \xi^{(1)}\circ \vp_i(\tx')))\cdot[(n_1\xi^{(1)}+n_2\xi^{(2)})
      (n_1(\frac{\pa u_\tau}{\pa \nu}\cdot\xi^{(1)})
          +n_2(\frac{\pa u_\tau}{\pa \nu}\cdot\xi^{(2)}))\ek
\hspace{2.4cm}  + (-n_2\xi^{(1)}+n_1\xi^{(2)})
   (-n_2(\frac{\pa u_\tau}{\pa \nu}\cdot\xi^{(1)})+n_1(\frac{\pa u_\tau}{\pa \nu}\cdot\xi^{(2)}))
      ]\ek
\hspace{2cm}  =  (n_1d_{11}+n_2d_{12})
      (n_1(\frac{\pa u_\tau}{\pa \nu}\cdot\xi^{(1)})+n_2(\frac{\pa u_\tau}{\pa \nu}\cdot\xi^{(2)}))\ek
   \hspace{2.4cm}           + (-n_2d_{11}+n_1d_{12})
      (-n_2(\frac{\pa u_\tau}{\pa \nu}\cdot\xi^{(1)})+n_1(\frac{\pa u_\tau}{\pa \nu}\cdot\xi^{(2)}))\ek
      \hspace{2cm} =d_{11}(\frac{\pa u_\tau}{\pa \nu}\cdot\xi^{(1)})+d_{12}(\frac{\pa u_\tau}{\pa \nu}\cdot\xi^{(2)}).
\end{array} $$
In the same way, one can check that the second component of $\Big[ C^{bl}(\Xi):
       \Big(\begin{array}{c}
         \la^{(1)}\frac{\pa u_\tau}{\pa\nu}\cdot \la^{(1)}\\
         \la^{(2)}\frac{\pa u_\tau}{\pa\nu}\cdot \la^{(2)}
         \end{array}
      \Big)\Big]$ is equal to
$d_{21}(\frac{\pa u_\tau}{\pa \nu}\cdot\xi^{(1)})+d_{22}(\frac{\pa u_\tau}{\pa \nu}\cdot\xi^{(2)})$.
Thus $c^{bl}\frac{\pa u_\tau}{\pa\nu}$ and hence $\bar\eta^\ve$ are
 irrespective of the choice of tangential vectors.

Next, in order to get the conclusion for $\eta^\ve$, it is enough to prove
$$\sum_{l=1}^2\beta_i^{\ve, \la^{(l)}(x')}(\frac{\widetilde{u_\tau}}{\pa\nu}\cdot\la^{(l)})
 =\sum_{l=1}^2\beta_i^{\ve, \xi^{(l)}(x')}(\frac{\widetilde{u_\tau}}{\pa\nu}\cdot\xi^{(l)})$$
for all $i=1,\ldots,N$ in view of the construction of $\eta^\ve$
(see \eq{3.38}, \eq{3.51}, \eq{3.35n}, \eq{3.33} and \eq{3.24-1}).
This equality follows directly by inserting $\la^{(1)}=n_1\xi^{(1)}+n_2\xi^{(2)}, \la^{(2)}=-n_2\xi^{(1)}+n_1\xi^{(2)}$
in view of the linearity of $\beta_i^{\ve, \la}$ w.r.t. $\la$ and $n_1^2+n_2^2=1$.

Thus, the proof of the lemma is complete.
\hfill $\Box$\vspace{0.2cm}

\par
\begin{lem}
\label{L3.13}
  {\rm Assume for $q$ the same as in Lemma \ref{L3.11n}.
  Then, for the function $\bar\eta^\ve$ defined by \eq{3.37} it
  holds
$$\|\na\bar\eta^\ve\|_{L^2(\ome)}
\les \ve\|u\|_{W^{2,q}(\Om)}.$$}
\end{lem}
\textbf{Proof.}
 For a vector $v\in \R^3$, let $v_j$ denote the $j$-th component of $v$.
 For any $\vp\in {\cal D}(\Om)^3$, $j=1\sim 3$, using integration by parts
 we have
   $$\begin{array}{l}
   \lan \eta^\ve_j,\vp\ran_{{\cal D}'(\Om^\ve),{\cal D}(\Om^\ve)}
    =  -(\eta^\ve_j,\div\vp)_{L^2(\Om^\ve)}\ek
    =  -\ve(\eta_j,\div\vp)_{L^2(\Om)}-\ve ((\tilde\Psi c^{bl}\frac{\widetilde{\pa u_\tau}}{\pa\nu})_j,\div\vp)_{L^2(\Om^\ve\setminus\Om)}\ek
    =  \ve(\na\eta_j,\vp)_{L^2(\Om)}-\ve\int_{\pa\Om}\eta_j \vp_{\nu}\,dx
        +\ve(\na(\tilde\Psi c^{bl}\frac{\widetilde{\pa u_\tau}}{\pa\nu})_j,\vp)_{L^2(\Om^\ve\setminus\Om)}
        -\ve\int_{\pa(\Om^\ve\setminus\Om)} (\tilde\Psi c^{bl}\frac{\widetilde{\pa u_\tau}}{\pa\nu})_j\vp_{\nu}\,dx.
\end{array}$$
Here, in view of $\vp|_{\pa\Om^\ve}=0$ and boundary condition in \eq{3.35},
$$-\int_{\pa(\Om^\ve\setminus\Om)} (\tilde\Psi c^{bl}\frac{\widetilde{\pa u_\tau}}{\pa\nu})_j\vp_{\nu}\,dx
= \int_{\pa\Om} (\Psi c^{bl}\frac{\widetilde{\pa u_\tau}}{\pa\nu})_j\vp_{\nu}\,dx
= \int_{\pa\Om} (\eta_\tau)_j\vp_{\nu}\,dx
= \int_{\pa\Om} \eta_j\vp_{\nu}\,dx.
$$
Hence, by \eq{3.35nn}, \eq{3.44n} we have
$$\begin{array}{rcl}
\lan \na\eta^\ve_j,\vp\ran_{{\cal D}'(\Om^\ve),{\cal D}(\Om^\ve)}
&=&\ve|(\na\eta_j,\vp)_{L^2(\Om)}
 +(\na(\tilde\Psi c^{bl}\frac{\widetilde{\pa u_\tau}}{\pa\nu})_j,\vp)_{L^2(\Om^\ve\setminus\Om)}|
 \ek
&\leq &\ve(\|\na\eta\|_{L^2(\Om)}+ \|\na(\tilde\Psi c^{bl}\frac{\widetilde{\pa u_\tau}}{\pa\nu})\|_{L^2(\Om^\ve\setminus\Om)})\|\vp\|_{L^2(\Om)}\ek
&\les &\ve\|u\|_{W^{2,q}(\Om)}
 \end{array}$$
for all $j=1\sim 3$, yielding
$$\na\eta^\ve\in L^2(\Om^\ve),\quad
 \|\na\eta^\ve\|_{L^2(\Om^\ve)}\les \ve\|u\|_{W^{2,q}(\Om)}$$ by denseness argument.
Thus the lemma is proved.
\hfill $\Box$

\begin{lem}
\label{L3.11} {\rm Let $r\geq 2$, $r'=\frac{r}{r-1}$ and let
$|g(x',\tx_3)|\les e^{\alpha\tx_3/\ve}$ in the curvilinear
coordinate system $(x',\tx_3)$ in $\Ga_\da^\ve$. Then, there holds
the following:\par
 (i) If $h\in
W^{1,2}(\Om^\ve)$, $h|_{\pa\Om^\ve\cap\pa\Ga_\da^\ve}=0$, then
$$\|g h\|_{L^{r'}(\Ga_\da^\ve)}\les
\ve^{3/2-1/r}\|h\|_{W^{1,2}(\Ga_\da^\ve)}.$$

(ii) If $h\in H^2(\Om')\cap H^1_{0}(\Om')$ where $\Om'=\{x\in \Om:
d(x,\Ga)>M\ve\}$, then
$$\begin{array}{l}
\|g h\|_{L^{r'}(\Ga_\da^\ve\cap\Om')}\les
\ve^{2-1/r}\|h\|_{H^{2}(\Ga_\da^\ve\cap\Om')},\ek
 \|g\na h\|_{L^{r'}(\Ga_\da^\ve\cap\Om')}\les
\ve^{1-1/r}\|h\|_{H^{2}(\Ga_\da^\ve\cap\Om')}.
\end{array}$$
 }
\end{lem}
\textbf{Proof.} - {\it Proof of (i):}
\par
By (74) of \cite{NNM06}, one has
$$\int_\Ga|h(x',\tx_3)|^{r'}
\,dx'\les (M\ve-\tx_3)^{r'/2}\|\na h\|_{L^2(\Ga_\da^\ve)}^{r'}$$
leading to
$$\begin{array}{rcl}
\Big(\int_{\Ga_\da^\ve}|g h(x',\tx_3)|^{r'}\,dx'd\tx_3\Big)^{1/{r'}}
&\les & \Big(\int_{-\da}^{M\ve}e^{\alpha r'\tx_3/\ve} \int_\Ga
|h(x',\tx_3)|^{r'}\,dx'd\tx_3\Big)^{1/{r'}}\ek
 &\les& \Big(\int_{-\da}^{M\ve}e^{\alpha r'\tx_3/\ve}
(M\ve-\tx_3)^{r'/2}\,d\tx_3\Big)^{1/{r'}}\|\na
h\|_{L^2(\Ga_\da^\ve)}.
 \end{array}$$
Here,
$$\int_{-\da}^{M\ve}e^{\alpha r'\tx_3/\ve}
(M\ve-\tx_3)^{r'/2}\,d\tx_3\les
\ve^{{r'}/2+1}\int_0^{\infty}e^{-\alpha r'y}y^{{r'}/2}\,dy,$$ which
completes the proof of (i).\par
 - {\it Proof of (ii):}
 \par
See page 498. of \cite{NNM06}.
 \hfill $\Box$ 
\begin{lem}
\label{L3.12} {\rm Let $h\in L^r(\Ga_\da^\ve), r \geq 1,$ satisfy
$|h(x)|\les e^{-\al d(x,\,\,\Ga)/\ve}, x\in \Ga_\da^\ve$.
 Then,
$$\|h\|_{L^r(\Ga_\da^\ve)}\les \ve^{1/r}.$$}
\end{lem}
\begin{lem}
\label{L3.14} {\rm Assume for $q$ the same as in Lemma
\ref{L3.11n}. Then, for function $\eta^\ve$ defined by \eq{3.38} it
holds
\begin{equation}
 \label{E3.60}
 (\na\eta^\ve,\na\varphi)_{\ome}=
-\int_{\Ga}\Psi(\frac{\pa u}{\pa\nu}-p\nu)\cdot\varphi\,ds +O(\ve)
  (\|u\|_{W^{2,q}(\Om)}+\|p\|_{W^{1,q}(\Om)})\|\na\varphi\|_{L^2(\ome)}
\end{equation}
for all $\vp\in H^1_{0,\si}(\Om^\ve)$.
}
\end{lem}
{\bf Proof.} Given any $\vp\in H^1_{0,\si}(\ome)$, one has
\begin{equation}
\label{E3.53}
\begin{array}{rcl}
  (\na\eta^\ve,\na\varphi)_{\ome}&=& \ve\sum_{l=1}^3\int_{\Ga_\da^\ve\setminus\Ga}\na
  (\beta^{\ve,l}\chi_l)\cdot\na\varphi\,dx\ek
&=& \ve\sum_{l=1}^3\int_{\Ga_\da^\ve\setminus\Ga}\chi_l\na
\beta^{\ve,l}\cdot\na\varphi\,dx
  +\ve\sum_{l=1}^3\int_{\Ga_\da^\ve\setminus\Ga}
  \beta^{\ve,l}\cdot(\na\chi_l\cdot\na\varphi)\,dx\ek
 & =:&(i)+(ii),
\end{array}
\end{equation}
where (ii) is estimated by
\begin{equation}
\label{E3.54}
\begin{array}{rcl}
 |(ii)|&\les& \ve\di\sup_{1\leq l\leq 3}\|\beta^{\ve,l}\na\chi_l\|_2
 \|\na\varphi\|_{L^2(\ome)}\ek
 &\les& \ve\di\sup_{1\leq l\leq
3}\|\beta^{\ve,l}\|_{L^{\infty}(\Ga_\da^\ve)}\|\na\chi_l\|_2
\|\na\varphi\|_{L^2(\ome)}\ek
 &\les&\ve(\|u\|_{W^{2,q}(\Om)}+\|p\|_{W^{1,q}(\Om)})\|\na\varphi\|_{L^2(\ome)}
\end{array}
\end{equation}
using  \eq{3.44n} and \eq{3.4}. On the other hand, we have
\begin{equation}
\label{E3.55}
\begin{array}{rl}
 (i)&=
 \ve\sum_{l=1}^3\int_{\Ga_\da^\ve\setminus\Ga}\div(\chi_l\na
  \beta^{\ve,l}\cdot\varphi)\,dx
  -\ve\sum_{l=1}^3\int_{\Ga_\da^\ve\setminus\Ga}\div(\na
  \beta^{\ve,l}\chi_l)\cdot\varphi\,dx\ek
 &= -\ve\sum_{l=1}^3\Big(\int_{\Ga}\chi_l\big[\frac{\pa
  \beta^{\ve,l}}{\pa\nu}\big]_\Ga\cdot\varphi\,ds
  +\int_{\Ga_\da^\ve\setminus\Ga}\chi_l\Da
  \beta^{\ve,l}\cdot\varphi\,dx\Big) -\ve\sum_{l=1}^3\int_{\Ga_\da^\ve\setminus\Ga}
   (\na\beta^{\ve,l}\cdot\na\chi_l)\cdot\varphi\,dx,
  \end{array}
\end{equation}
where
\begin{equation}
\label{E3.55n}
\begin{array}{l}
\ve|\int_{\Ga_\da^\ve\setminus\Ga}\na
  \beta^{\ve,l}\cdot\na\chi_l\varphi\,dx| \les
  \|\na\chi_l\|_{L^q(\Ga_\da^\ve)}\|\ve\na\beta^{\ve,l}\varphi\|_{L^{q'}(\Ga_\da^\ve)}\ek
\hspace{2cm}\les \|\na\chi_l\|_{L^q(\Ga_\da^\ve)}(\|\ve\na\beta^{\ve,l}\varphi\|_{L^{q'}(\Ga_\da^\ve\cap\Om)}
       +\|\ve\na\beta^{\ve,l}\|_{L^{2q/(q-2)}(\Ga_\da^\ve\setminus \Om)}
        \|\varphi\|_{L^{2}(\Ga_\da^\ve\setminus \Om)})\ek
\hspace{2cm} \les
\ve(\|u\|_{W^{2,q}(\Om)}+\|p\|_{W^{1,q}(\Om)})\|\na\varphi\|_{L^2(\ome)}
\end{array}
\end{equation}
by Lemma \ref{L3.11} (i), \eq{3.44n}, Lemma \ref{L3.9} (i) and Poincar\'e's inequality.
 On the other hand, by Lemma \ref{L3.9} (iv) the first term in
 the right-hand side of \eq{3.55} is expanded as follows:
\begin{equation}
 \label{E3.58}
 \begin{array}{l}
-\ve\sum_{l=1}^3\big(\int_{\Ga}\chi_l\big[\frac{\pa
  \beta^{\ve,l}}{\pa\nu}\big]_\Ga\cdot\varphi\,ds
  +\int_{\Ga_\da^\ve\setminus\Ga}\chi_l\Da
  \beta^{\ve,l}\cdot\varphi\,dx\big)\ek
 \hspace{1cm} =-\sum_{l=1}^3\int_{\Ga}\chi_l([\om^{\ve,l}\nu]_\Ga+\la^{(l)})\cdot\varphi\,ds
-\ve\sum_{l=1}^3\int_{\Ga_\da^\ve\setminus\Ga}
   \chi_l \Da\beta^{\ve,l}\cdot\varphi\,dx\ek
\hspace{1cm}=-\sum_{l=1}^3\int_{\Ga}\chi_l([\om^{\ve,l}]_\Ga\vp_\nu\,ds-\int_\Ga\chi\cdot\varphi\,ds
-\ve\sum_{l=1}^3\int_{\Ga_\da^\ve\setminus\Ga}
   \chi_l \Da\beta^{\ve,l}\cdot\varphi\,dx\ek
\hspace{1cm}=-\int_\Ga\chi\cdot\varphi\,ds
+\sum_{l=1}^3\int_{\Ga_\da^\ve\setminus\Ga}
  (\div(\om^{\ve,l}\chi_l\vp)-\ve\chi_l \Da\beta^{\ve,l}\cdot\varphi)\,dx\ek
\hspace{1cm}  =-\int_{\Ga}\Psi(\frac{\pa u}{\pa\nu}-p\nu)\cdot\varphi\,ds
+\int_{\Ga\setminus\Ga'}\Psi\frac{\pa
u_\nu}{\pa\nu}\varphi_\nu\,ds\ek
\hspace{1.5cm}-\sum_{l=1}^3\int_{\Ga_\da^\ve\setminus\Ga}\chi_l(\ve\Da
  \beta^{\ve,l}-\na\om^{\ve,l})\cdot\varphi\,dx
+\sum_{l=1}^3\int_{\Ga_\da^\ve\setminus\Ga}
\om^{\ve,l}\na\chi_l\cdot\varphi\,dx.
\end{array}
\end{equation}
Here we used that
$$\frac{\pa u_\tau}{\pa\nu}\cdot\vp=(\frac{\pa
u}{\pa\nu}-\frac{\pa u_\nu}{\pa\nu}\nu-u_\nu\nu)\cdot\vp=\frac{\pa
u}{\pa\nu}\cdot\vp$$
 since
$\frac{\pa u_\nu}{\pa\nu}|_{\Ga'}=0$ in view of $\div u=0,
u|_{\Ga'}=0$. The last two terms in the right-hand side of \eq{3.58}
are shown to be equal to
$$O(\ve)(\|u\|_{W^{2,q}(\Om)}+\|p\|_{W^{1,q}(\Om)})\|\na\varphi\|_{L^2(\ome)}$$
by Lemma \ref{L3.9} (i), (ii), Lemma \ref{L3.11} (i) and \eq{3.44n}.

 Let
$\Ga\setminus\Ga'$ be nontrivial ($q>3$ in this case). Since the width of
two-dimensional annular disc $\Ga\setminus\Ga'$ is $O(\ve)$ and $\vp=0$
on its outer boundary $\bar\Ga\cap\Ga_1$, we get by Poincar\'e's
inequality $\|\vp\|_{L^2(\Ga\setminus\Ga')}\les
\ve\|\frac{\pa\vp}{\pa\tau}\|_{L^2(\Ga\setminus\Ga')}$. Then, by
complex interpolation we get that
\begin{equation}
\label{E3.62nn} \|\vp\|_{L^2(\Ga\setminus\Ga')}\les
\ve^{1/2}\|\vp\|_{H^{1/2}(\Ga\setminus\Ga')}\les
\ve^{1/2}\|\na\vp\|_{L^2(\Om)}.
\end{equation}
Consequently, by Sobolev embedding theorem in view of $q>3$
 we have
 $$\begin{array}{rcl} \big|\int_{\Ga\setminus\Ga'}\Psi\frac{\pa
u_\nu}{\pa\nu}\varphi_\nu\,ds\big| &\leq & \big\|\frac{\pa
u}{\pa\nu}\big\|_{L^2(\Ga\setminus\Ga')}
\|\varphi\|_{L^2(\Ga\setminus\Ga')}\ek
 &\les &\big\|\frac{\pa
 u}{\pa\nu}\big\|_{L^\infty(\Ga\setminus\Ga')}|\Ga\setminus\Ga'|^{1/2}\cdot\ve^{1/2}
\|\na\varphi\|_{L^2(\Om)}\ek
&\les&\ve\|u\|_{W^{2,q}(\Om)}\|\na\varphi\|_{L^2(\ome)}.
\end{array}$$
The proof of the lemma is complete. \hfill $\Box$

\begin{lem}
\label{L3.13n}
 {\rm Assume for $q$ as in Lemma \ref{L3.11n}.
 Then the inhomogeneous boundary value problem
\begin{equation}
\label{E3.43}
\begin{array}{rccl}
 -\Da w^\ve+\na r^\ve&= & 0 &\quad\text{in }\ome,\ek
 \div w^\ve &=& -\div(\bar\eta^\ve+\eta^\ve) &\quad\text{in }\ome,\ek
 w^\ve &= & 0 &\quad\text{on }\pa\ome,
\end{array}
\end{equation}
has a unique weak solution $\{w^\ve,r^\ve\}\in H^1_0(\Om^\ve)\ti L^2_{(m)}(\ome)$ such that
\begin{equation}
\label{E3.45}
 \|\na w^\ve\|_{L^2(\ome)}+ \|r^\ve\|_{L^2(\ome)}\les
\ve(\|u\|_{W^{2,q}(\Om)}+\|p\|_{W^{1,q}(\Om)}).
\end{equation}
}
 \end{lem}
\textbf{Proof.} First of all, we remark that for every $g\in L^2_{(m)}(\Om^\ve)$
the divergence problem
$$\div \psi = g\quad\text{in }\Om^\ve, \quad \psi|_{\pa\Om^\ve}=0,$$
has a solution $\psi\in W^{1,2}_0(\Om^\ve)$ satisfying the estimate
$$\|\psi\|_{W^{1,2}_0(\Om^\ve)}\leq C \|g\|_{L^{2}(\Om^\ve)},$$
  where the constant $C$ is independent of $\ve$.
This fact follows by Appendix, Lemma A.1 and Lemma A.2 using
the assumption \eq{1.2} on $\Om^\ve$  that
$\Om^\ve$ can be expressed by sum of several rough domains $G^{(j)\ve}$ where
$G^{(j)\ve}, j=1,\ldots,m,$ is again a sum of one "main" macroscopic star-shaped domain
and many microscopic $O(\ve)$-size star-shaped domains, i.e.,
  $$G^{(j)\ve}=G_0^{(j)}\cup \bigcup_{k=1}^{m_j}G_k^{(j)},\quad
G_0^{(j)}\cap G_k^{(j)}\neq \emptyset, G_k^{(j)}\cap G_l^{(j)}=\emptyset,
k\neq l, k,l=1,\ldots, m_j.$$

Let $\psi  \in W^{1,2}_0(\Om^\ve)$ be such that $\div \psi=-\div(\bar\eta^\ve+\eta^\ve)$
and $\|\psi\|_{W^{1,2}_0(\Om^\ve)}\les \|\div(\bar\eta^\ve+\eta^\ve)\|_{L^2(\Om^\ve)}$.
 Then, it is standard to show the
existence of unique weak solution $\{w^\ve,r^\ve\}$ to the problem
 \eq{3.43} such that
 $$\|\na w^\ve\|_{L^2(\ome)}+ \|r^\ve\|_{L^2(\ome)}\les \|\na\psi\|_{L^2(\Om^\ve)}\les
\|\div (\bar\eta^\ve+\eta^\ve)\|_{L^2(\ome)}.$$
By the way,
 we get
$$\|\div \bar\eta^\ve\|_{L^2(\ome)} \leq
 \ve\|\tilde\Psi c^{bl}\frac{\widetilde{\pa u_\tau}}{\pa\nu}\|_{H^1(\ome\setminus\Om)}
 \les  \ve(\|u\|_{W^{2,q}(\Om)}+\|p\|_{W^{1,q}(\Om)})
 $$
from \eq{3.34}, \eq{3.44n}. Moreover, we have
\begin{equation}
\label{E3.44}
\begin{array}{l} \|\div \eta^\ve\|_{L^2(\ome)} \leq
\ve\sum_{l=1}^3(\|\div \beta^{\ve,l}\chi_l\|_{L^2(\Ga_\da^\ve)}+
                \|\beta^{\ve,l}\cdot\na\chi_l\|_{L^2(\Ga_\da^\ve)})\ek
\hspace{1cm}\leq
\ve\sum_{l=1}^3(\|\div\beta^{\ve,l}\|_{L^{\infty}(\Ga_\da^\ve\cap\Om)}+
  \|\div\beta^{\ve,l}\|_{L^{2q/(q-2)}(\Ga_\da^\ve\setminus\Om)}+
\|\beta^{\ve,l}\|_{L^\infty(\Ga_\da^\ve)})\|\chi_l\|_{W^{1,q}(\Ga_\da^\ve)}\ek
\hspace{1cm}\les \ve(\|u\|_{W^{2,q}(\Om)}+\|p\|_{W^{1,q}(\Om)})
\end{array}
\end{equation}
from \eq{3.38}, Lemma \ref{L3.9} (iii) and \eq{3.44n}.

Thus, the proof comes to end.\hfill $\Box$

Now we can prove the following theorem on the error estimates of
first order approximation for $u^\ve$.
\begin{theo}
\label{T3.14} {\rm Assume for $q$ the same as in Lemma
\ref{L3.11n}.
 Let $f\in L^q(\Om^\ve)$, $\psi\in W^{2-1/q,q}(\pa\ome)$,
 $\supp\psi\subset\Ga_1$ and let
 $u\in W^{2,q}(\Om_0)$ be the solution to \eq{3.3}. Then, the estimates
\begin{equation}
\label{E3.39}
\|\na(u^\ve-(\tilde{u}+\bar\eta^\ve+\eta^\ve))\|_{L^2(\ome)}\les
   \ve(\|f\|_{L^q(\ome)}+\|\psi\|_{W^{2-1/q,q}(\Ga_1)})
\end{equation}
and
\begin{equation}
\label{E3.40}
\|u^\ve-(\tilde{u}+\bar\eta^\ve+\eta^\ve)\|_{L^2(\ome)}\les
   \ve^{3/2}(\|f\|_{L^q(\ome)}+\|\psi\|_{W^{2-1/q,q}(\Ga_1)})
\end{equation}
hold true.}
\end{theo}
\textbf{Proof.} Let
\begin{equation}
\label{E3.47}
 v^\ve:=u^\ve-(\tilde{u}+\bar\eta^\ve+\eta^\ve+w^\ve),
\end{equation}
where $w^\ve$ is the solution to the system \eq{3.43} and
$\bar\eta^\ve$, $\eta^\ve$ are defined by \eq{3.35}, \eq{3.38},
respectively. Then, we have $v^\ve\in H^1_{0,\si}(\ome)$ since $\div v^\ve =
0$ in $\ome$ and $v^\ve =0$ on $\pa\ome$.
\par
For any $\varphi\in H^1_{0,\si}(\ome)$ we have
\begin{equation}
\label{E3.48}
\begin{array}{rl}
(\na v^\ve,\na \varphi)_{\ome}
 =&(\na(u^\ve-(\tilde{u}+\bar\eta^\ve+\eta^\ve+w^\ve)),
  \na\varphi)_{\ome}\ek
  =& -\int_\Ga(\frac{\pa u}{\pa\nu}-p\nu)\cdot\varphi\,ds
  -(\na\eta^\ve,\na\varphi)_{\ome}
  -(\na\bar\eta^\ve, \na\varphi)_{\ome}\ek &
  -(\na w^\ve, \na\varphi)_{\ome}+(f,\varphi)_{\ome\setminus\Om}
   -(\na u, \na\varphi)_{\Om_0\setminus\Om}.
\end{array}
\end{equation}
By Lemma \ref{L3.14}, Lemma \ref{L3.13}, \eq{3.45} and
\eq{3.4}, the sum of the first four terms in the right-hand side of
\eq{3.48} is equal to
$$\int_\Ga(\Psi-1)(\frac{\pa u}{\pa\nu}-p\nu)\cdot\varphi\,dx'
+O(\ve)(\|f\|_{L^q(\ome)}+
\|\psi\|_{W^{1-1/q,q}(\Ga_1)})\|\na\varphi\|_{L^2(\ome)}.$$
Moreover, by Poincar\'e's inequality the fifth term in the
right-hand side of \eq{3.48} is estimated by
$$ \begin{array}{rcl}
 |(f,\varphi)_{\ome\setminus\Om}|\leq
 \|f\|_{L^2(\ome\setminus\Om)}\|\varphi\|_{L^2(\ome\setminus\Om)}
 \les \ve
  \|f\|_{L^2(\ome\setminus\Om)}\|\na\varphi\|_{L^2(\ome)},
 \end{array}
$$
and  the sixth term by
$$ \begin{array}{rcl}
 \|\na u\|_{L^2(\Om_0\setminus\Om)}&\leq& \|\na
 u\|_{L^6(\Om_0\setminus\Om)} |\Om_0\setminus\Om|^{1/3}\ek
    &\les& \ve \|
  u\|_{W^{2,2}(\Om_0)}\ek
&\les&\ve(\|f\|_{L^2(\ome)}+\|\psi\|_{W^{3/2,2}(\Ga_1)})
 \end{array}
$$
using  Sobolev embedding theorem and $|\Om_0\setminus\Om|\leq O(\ve^3)$.
Therefore, if we prove
\begin{equation}
\label{E3.62n} \int_\Ga(\Psi-1)(\frac{\pa
u}{\pa\nu}-p\nu)\cdot\varphi\,dx' \leq
\ve(\|f\|_{L^q(\ome)}+\|\psi\|_{W^{2-1/q,q}(\Ga_1)}),
 \end{equation}
then we get
$$|(\na v^\ve,\na\varphi)_{\ome}|\les  \ve
  (\|f\|_{L^q(\ome)}+\|\psi\|_{W^{2-1/q,q}(\Ga_1)})\|\na\varphi\|_{L^2(\ome)},
  \quad\forall\varphi\in H^1_{0,\si}(\ome),$$
and hence \eq{3.39} by Lemma \ref{L3.13n}.
We need to consider only the case where $\Ga$ and $\Ga_1$ are adjacent. Note that
 $q>3$ in this case.
It follows from the construction of $\Psi$ and \eq{3.62nn} that
$$\begin{array}{rcl}
|\int_\Ga(\Psi-1)(\frac{\pa u}{\pa\nu}-p\nu)\cdot\varphi\,dx'|
 &\les & \|\frac{\pa
u}{\pa\nu}-p\nu\|_{L^2(\Ga\setminus\Ga')}\|\vp\|_{L^2(\Ga\setminus\Ga')}\ek
 &\les & \ve^{1/2}\|\frac{\pa
u}{\pa\nu}-p\nu\|_{L^2(\Ga\setminus\Ga')}\|\na\vp\|_{L^2(\Om)} ,
\end{array}
$$
where
$$\|\frac{\pa
u}{\pa\nu}-p\nu\|_{L^2(\Ga\setminus\Ga')}\leq \|\frac{\pa
u}{\pa\nu}-p\nu\|_{L^\infty(\Ga)}|\Ga\setminus\Ga'|^{1/2}\les
\ve^{1/2}(\|u\|_{W^{2,q}(\Om)}+\|p\|_{W^{1,q}(\Om)}).
$$
Hence,
we get \eq{3.62n} due to \eq{3.4} and \eq{3.39} is proved.
\par
Next, in order to prove \eq{3.40}, we use the idea of \cite{NNM06}.
Let $$z:=u^\ve-(\tilde{u}+\bar\eta^\ve+\eta^\ve),\quad \Om':=\{x\in\Om:
\rho(x)>M\ve\}.$$
 Then, by Poincar\'e's inequality and already proved \eq{3.39}
 we have
\begin{equation}
\label{E3.61}
 \|z\|_{L^2(\ome\setminus\Om')}\les \ve\|\na z\|_{L^2(\ome\setminus\Om')}
\les \ve^{2}(\|f\|_{L^q(\ome)}+\|\psi\|_{W^{2-1/q,q}(\Ga_1)}).
\end{equation}
Therefore, for the proof of \eq{3.40} we only need to estimate
$\|z\|_{L^2(\Om')}$. Let $w\in H^2(\Om')\cap H^1_{0,\si}(\Om')$
be the unique solution to the Stokes problem
\begin{equation}
\label{E3.62}
\begin{array}{rl}
 -\Da w+\na s=z &  \quad\text{in }\Om',\ek
 \div w = 0 &\quad \text{in }\Om',\ek
 w =0 & \quad\text{on }\pa\Om'.
\end{array}
\end{equation}
Then, one has $\|w\|_{H^2(\Om')}+\|s\|_{H^1(\Om')}\les \|z\|_{L^2(\Om')}$ and
\begin{equation}
\label{E3.63}
 \|z\|^2_{L^2(\Om')}=\int_{\Om'}(-\Da w+\na s)\cdot z\,dx
=\int_{\Om'}(\na w\cdot \na z - s \div z)\,dx
 -\int_{\pa\Om'}(\frac{\pa w}{\pa\nu}-s\nu)\cdot z\,ds.
\end{equation}
Note that the estimate \eq{2.5n} in Lemma \ref{L2.3} holds for
$\Om'$ as well. Hence, it follows from $z|_{\pa\Om^\ve}=0$ and
\eq{3.39} that
$$\|z\|_{L^2(\pa\Om')}\les \ve^{1/2}\|\na z\|_{L^2(\ome\setminus \Om')}
\les \ve^{3/2}(\|f\|_{L^q(\ome)}+\|\psi\|_{W^{2-1/q,q}(\Ga_1)}).$$
Therefore the second term in the right-hand side of \eq{3.63} is
estimated by
$$
\begin{array}{rcl}
 \big|\int_{\pa\Om'}(\frac{\pa w}{\pa\nu}-s\nu)\cdot z\,ds\big|
  &\les& \|\na w-s\|_{H^1(\Om')}\|z\|_{L^2(\pa\Om')}\ek
  &\les& \|z\|_{L^2(\Om')}\|z\|_{L^2(\pa\Om')}\ek
  &\les&
  \ve^{3/2}\|z\|_{L^2(\Om')}(\|f\|_{L^q(\ome)}+\|\psi\|_{W^{2-1/q,q}(\Ga_1)}).
 \end{array}
 $$

 Let us get estimate of the first term in the right-hand side of
\eq{3.63}. Obviously, we have
$$
\int_{\Om'}(\na w\cdot \na z - s \div z)\,dx
  =-\int_{\Om'}(\na w\cdot \na\eta^\ve-s \div \eta^\ve)\,dx.
$$
where
\begin{equation}
\label{E3.63n}
\begin{array}{l}
\int_{\Om'}\na w\cdot\na\eta^\ve\,dx = \ve\sum_{l=1}^3
\int_{\Om'}\na w\cdot\na(\beta^{\ve,l}\chi_l)\,dx\ek
  \hspace{0.5cm} =\ve\sum_{l=1}^3 \int_{\Om'}\big((\na w\cdot\na\chi_l)\cdot\beta^{\ve,l}
   +\na w\cdot\na\beta^{\ve,l}\chi_l \big)\,dx\ek
  \hspace{0.5cm}= \sum_{l=1}^3 \big(\ve\int_{\Om'}(\na w\cdot\na\chi_l)\cdot\beta^{\ve,l}\,dx\ek
  \hspace{0.6cm}+\int_{\Om'} w\cdot
  (-\ve\Da\beta^{\ve,l}\chi_l-\ve\na\beta^{\ve,l}\cdot\na\chi_l)\,dx
   +\int_{\Om'} w \cdot\na(\om^{\ve,l}\chi_l)\,dx   \big)\ek
  \hspace{0.5cm}= \ve\sum_{l=1}^3 \int_{\Om'}(\na
w\cdot\na\chi_l)\cdot\beta^{\ve,l}\,dx \ek
 \hspace{0.6cm} +\sum_{l=1}^3
\int_{\Om'}\big(w\cdot(-\ve\Da\beta^{\ve,l}+\na\om^{\ve,l})\chi_l
-(w\cdot\na\chi_l)\cdot(\ve\na\beta^{\ve,l})
   + w \cdot\na\chi_l\om^{\ve,l}\big)\,dx.
\end{array}
\end{equation}
By Lemma \ref{L3.9}, $\beta^{\ve,l}$,
$-\ve\Da\beta^{\ve,l}+\na\om^{\ve,l}$, $\ve\na\beta^{\ve,l}$,
$\om^{\ve,l}$ in the right-hand side of \eq{3.63n} decays at the
rate of $e^{-\alpha \rho(x)/\ve}$ near $\Ga$. Therefore, by Lemma
\ref{L3.11} (ii), \eq{3.44n} and \eq{3.4} we have
$$\begin{array}{rcl}
|\int_{\Om'}\na
w\cdot\na\eta^\ve\,dx|
   &\les&\ve^{3/2}\sum_{l=1}^3(\|\na\chi_l\|_2+\|\chi_l\|_2)\|w\|_{H^2(\Om')}\ek
   &\les&\ve^{3/2}(\|u\|_{W^{2,q}(\Om)}+\|p\|_{W^{1,q}(\Om)})\|z\|_{L^2(\Om')}\ek
   &\les&\ve^{3/2}(\|f\|_{L^q(\ome)}+\|\psi\|_{W^{2-1/q,q}(\Ga_1)})\|z\|_{L^2(\Om')}.
\end{array}
$$
In the same way,  using decay estimate of $\div
\beta^{\ve,l},l=1\sim 3$, given by Lemma \ref{L3.9} (iii), the
integral of $s\div\eta^\ve$ can be estimated  with the same order of $O(\ve^{3/2})$.

Thus we have
$$\|z\|_{L^2(\Om')}\les\ve^{3/2}(\|f\|_{L^q(\ome)}+\|\psi\|_{W^{2-1/q,q}(\Ga_1)}),$$
and hence \eq{3.40}.
 \hfill $\Box$ \vspace{0.2cm}
\par
Now, let us construct an effective Navier
wall-law for the Stokes system \eq{3.1} as follows:
\begin{equation}
\label{E3.66}
\begin{array}{rccl}
 -\Da u^{eff}+\na p^{eff}&=&f &  \quad\text{in }\Om,\ek
 \div u^{eff}& =& 0 &\quad\text{in }\Om,\ek
 u_{\tau}^{eff} &=&\ve \Psi c^{bl}\frac{\pa u_{\tau}^{eff}}{\pa\nu} &
\quad \text{on }\Ga,\ek
 u_\nu^{eff} &=&0 & \quad\text{on }\Ga,\ek
 u^{eff} &=&\psi & \quad \text{on }\Ga_1.
\end{array}
\end{equation}

\begin{rem}
{\rm
The Navier wall-law of \eq{3.66} is irrespective of the
choice of curvilinear systems of orthogonal tangential vectors
on $\Ga$ due to Lemma \ref{L3.15}.
}
\end{rem}

Since the matrix $c^{bl}(x')$ is negatively definite and
$\Psi(x')\geq 0$ for all $x'\in\Ga$, the problem \eq{3.66} is
well-posed and has a weak solution $u^{eff}\in H^1(\Om)$
by Lax-Milgram's lemma.
\begin{theo}
\label{T3.15} {\rm Assume for $q$ the same as in Lemma
\ref{L3.11n}. Let $u$ be the solution to \eq{3.3}, and let
$\bar\eta^\ve$ be defined by \eq{3.37}.
 Then,
$$\|u^{eff}-u-\bar\eta^\ve\|_{H^1(\Om)}\les
\ve(\|f\|_{L^q(\ome)}+\|\psi\|_{W^{2-1/q,q}(\Ga_1)}),$$
$$\|u^{eff}-u-\bar\eta^\ve\|_{L^2(\Om)}\les
\ve^{3/2}(\|f\|_{L^q(\ome)}+\|\psi\|_{W^{2-1/q,q}(\Ga_1)}).$$
}
\end{theo}
\textbf{Proof.} Let $v:=u^{eff}-u-\bar\eta^\ve$. Then, $v$ solves the
system
\begin{equation}
\label{E3.68}
\begin{array}{rcll}
 -\Da v+\na s &=&0 &  x\in\Om,\ek
 \div v &=& 0 &x\in\Om,\ek
 v_{\tau} &=&\phi&
          x'\in\Ga,\ek
          v_{\nu}&=&-u_\nu\chi_{\Ga\setminus\Ga'} &  x'\in\Ga,\ek
 v&=&0 &  x'\in\Ga_1,
\end{array}
\end{equation}
where
 $\phi=\ve \Psi(x')c^{bl}(x')(\frac{\pa v_{\tau}}{\pa\nu}+\frac{\pa
 \bar\eta^\ve_{\tau}}{\pa\nu})-u_\tau\chi_{\Ga\setminus\Ga'}$,
 $\chi_{\Ga\setminus\Ga'}$ is the characteristic function of $\Ga\setminus\Ga'$.

For the associate pressure $s$ we may assume without loss of
generality that $s\in L^2_{(m)}(\Om)$.
Then,
 $$\|s\|_{L^2_{(m)}(\Om)} \les
\|\na v\|_{L^2(\Om)}.$$
 In fact, given any $h\in L^2_{(m)}(\Om)$ there is some $\varphi\in
H^1_0(\Om)$ such that $\div \varphi =h$, $\|\varphi\|_{H^1_0(\Om)}\leq
c(\Om)\|h\|_{L^2_{(m)}(\Om)}$. Hence,
$$(s,h)_\Om=(s,\div\varphi)_\Om=(-\na s,\varphi)_\Om=(\na v,\na\varphi)_\Om$$
and
$|(s,h)|\leq \|\na v\|_2\|\na\varphi\|_2\les \|\na
v\|_2\|h\|_{L^2_{(m)}(\Om)}$ implying
$\|s\|_{L^2_{(m)}(\Om)}\les \|\na v\|_2$.

Since $\na v, s\in L^2(\Om)$ and $\div(\na v-sI)=0$, we get that
\begin{equation}
\label{E3.70}
 (\frac{\pa
v}{\pa\nu}-s\nu)|_{\Ga}\in H^{-\ha}(\Ga),\quad \|\frac{\pa
v}{\pa\nu}-s\nu\|_{H^{-\ha}(\Ga)}\les \|\na v-sI\|_{L^2(\Om)}\les
\|\na v\|_{L^2(\Om)},
\end{equation}
see \cite{Ga94-1}, Ch. 3, Theorem 2.2; cf. also \cite{Te77}.
 By the same reasoning, for  the solution $\{\eta,\zeta\}$ to
\eq{3.35} we have
\begin{equation}
\label{E3.71}
 (\frac{\pa \eta}{\pa\nu}-\zeta\nu)|_\Ga\in H^{-\ha}(\Ga), \quad
  \|\frac{\pa \eta}{\pa\nu}-\zeta\nu\|_{H^{-\ha}(\Ga)}\les \|\na \eta-\zeta I\|_{L^2(\Om)}\les
\|\na \eta\|_{L^2(\Om)}.
\end{equation}
Since the matrix $\Psi(x')c^{bl}(x')$  for any $x'\in\Ga$ is
invertible, it follows from the boundary condition of \eq{3.68} that
 $$\frac{\pa v_{\tau}}{\pa\nu}=
 (\ve\Psi c^{bl})^{-1}(v_\tau+u_\tau\chi_{\Ga\setminus\Ga'})
  -\frac{\pa \bar\eta_\tau^\ve}{\pa\nu}\quad \text{on }\Ga.$$
Hence,
by testing \eq{3.68} with $v$ in view of \eq{3.70} and negativity of $c^{bl}$, we have
\begin{equation}
\label{E3.68n}
\begin{array}{rl}
 \|\na v\|_{L^2(\Om)}^2 =&
  \lan v,\frac{\pa v}{\pa\nu}-s\nu\ran_{H^{\ha}(\Ga),H^{-\ha}(\Ga)}\ek
 =&\lan v_\tau,\frac{\pa v_\tau}{\pa\nu}\ran_{H^{\ha}(\Ga),H^{-\ha}(\Ga)}
  +\lan v_\nu,\frac{\pa v_\nu}{\pa\nu}-s+\frac{\pa v_\tau}{\pa\nu}\cdot\nu\ran_{H^{\ha}(\Ga),H^{-\ha}(\Ga)}\ek
   =&\lan v_\tau,\frac{\pa v_\tau}{\pa\nu}\ran_{H^{\ha}(\Ga),H^{-\ha}(\Ga)}
  -\lan u_\nu,\frac{\pa v_\nu}{\pa\nu}-s+\frac{\pa v_\tau}{\pa\nu}\cdot\nu\ran_{H^{\ha}(\Ga\setminus\Ga'),H^{-\ha}(\Ga\setminus\Ga')}\ek
  =& \int_{\Ga}v_\tau\cdot(\ve\Psi c^{bl})^{-1}(v_\tau
  +u_\tau\chi_{\Ga\setminus\Ga'})\,dx'-\lan v_\tau, \frac{\pa
\bar\eta_\tau^\ve}{\pa\nu}\ran_{H^{\ha}(\Ga), H^{-\ha}(\Ga)}\ek
&\hspace{0.5cm}-\lan u_\nu,\frac{\pa v_\nu}{\pa\nu}-s+\frac{\pa v_\tau}{\pa\nu}\cdot\nu\ran_{H^{\ha}(\Ga\setminus\Ga'),H^{-\ha}(\Ga\setminus\Ga')}
\ek
 \leq &\int_{\Ga\setminus\Ga'}v_\tau\cdot(\ve\Psi
 c^{bl})^{-1}u_\tau\,dx'
  -\lan v_\tau, \frac{\pa \bar\eta_\tau^\ve}{\pa\nu}\ran_{H^{\ha}(\Ga),
  H^{-\ha}(\Ga)}\ek
  &\hspace{0.5cm}-\lan u_\nu,\frac{\pa v_\nu}{\pa\nu}-s+\frac{\pa v_\tau}{\pa\nu}\cdot\nu\ran_{H^{\ha}(\Ga\setminus\Ga'),H^{-\ha}(\Ga\setminus\Ga')}.
  \end{array}
  \end{equation}
Since $u$ vanishes at the boundary of $\Om_0$ and the
thickness of the annular disc $\Om_0\setminus\Om$ is $O(\ve^2)$, it follows
that
\begin{equation}
\label{E3.69n} \|u\|_{L^{2}(\Ga\setminus\Ga')}\les\ve \|\na
u\|_{L^{2}(\Om_0\setminus\Om)}\les \ve\|\na
u\|_{L^{6}(\Om_0\setminus\Om)}|\Om_0\setminus\Om|^{1/3}
 \les \ve^2\|\na^2 u\|_{L^{2}(\Om_0)}
 \end{equation}
and, by trace theorem,
$$\|u\|_{H^{3/2}(\Ga\setminus\Ga')}\les \|\na^2 u\|_{L^{2}(\Om_0)}.$$
 Therefore,  it follows by complex interpolation
 $H^{\theta}(\Ga\setminus\Ga')=
 [L^2(\Ga\setminus\Ga'), H^{3/2}(\Ga\setminus\Ga')]_{\frac{2\theta}{3}}$
  for $1\leq \theta\leq 3/2$ that
\begin{equation}
\label{E3.74} \|u\|_{H^{\theta}(\Ga\setminus\Ga')}\leq
\|u\|^{{2\theta}/3}_{H^{3/2}(\Ga\setminus\Ga')}\|u\|^{1-{2\theta}/3}_{L^{2}(\Ga\setminus\Ga')}
 \les \ve^{2-{4\theta}/3}\|\na^2 u\|_{L^{2}(\Om_0)}.
\end{equation}
Consequently, the third term in the right-hand side of \eq{3.68n}
is estimated using \eq{3.70} as
\begin{equation}
\label{E3.68nnn}
\begin{array}{rcl}|\lan
u_\nu,\frac{\pa v_\nu}{\pa\nu}
 -s+\frac{\pa v_\tau}{\pa\nu}\cdot\nu\ran_{H^{\ha}(\Ga\setminus\Ga'),H^{-\ha}(\Ga\setminus\Ga')}|
  &\leq &\|u_\nu\|_{H^{\ha}(\Ga\setminus\Ga')}\|\frac{\pa v_\nu}{\pa\nu}
 -s+\frac{\pa v_\tau}{\pa\nu}\cdot\nu\|_{H^{-\ha}(\Ga\setminus\Ga')}\ek
&\les &\|u\|_{H^{\ha}(\Ga\setminus\Ga')}\|\na v\|_{L^2(\Om)}\ek
 &\les & \ve\|u\|_{W^{2,2}(\Om_0)}\|\na v\|_{L^2(\Om)}.
  \end{array}
  \end{equation}

On the other hand, due to the construction of $\Psi$, the first term
in the right-hand side of \eq{3.68n} is estimated as
\begin{equation}
\label{E3.68nn}
 \Big|\int_{\Ga\setminus\Ga'}v_\tau\cdot(\ve\Psi
 c^{bl})^{-1}u_\tau\,dx'\Big|
  \les
  \int_{\Ga\setminus\Ga'}|d(x',\bar\Ga\cap\Ga_1)^{-1}v_\tau\cdot u_\tau|\,dx'
  \end{equation}
Note that
  $$\|x_n^{-\ga} h\|_{L^2(\R^n_+)}\leq c(\ga)\|h\|_{H^\ga_0(\R^n_+)},\quad
 \forall h\in H^\ga_0(\R^n_+),\ga\in [0,1]\setminus\{\ha\},$$
 cf. \cite{LM72}, Ch.1, Theorem 11.3.
This inequality can be extended to the case where $\R^n_+$ is
replaced by $\Ga\setminus\Ga'$ using diffeomorphism  $\vp_i$ between
 $U_i\subset \R^2$ and $V_i\subset \Ga$ for $i=1,\ldots,N$.
 In particular, for $\ga\in [0,1]\setminus\{\ha\}$
$$\|d(x',\bar\Ga\cap\Ga_1)^{-\ga}z\|_{L^2(\Ga\setminus\Ga')}
\leq
 c(\ga,\Ga\setminus\Ga')\|z\|_{H^\ga_{0,\bar\Ga\cap\Ga_1}(\Ga\setminus\Ga')},
\quad \forall z\in H^\ga_{0,\bar\Ga\cap\Ga_1}(\Ga\setminus\Ga'),$$
where $H^\ga_{0,\bar\Ga\cap\Ga_1}(\Ga\setminus\Ga')$ is the closure
 in $H^\ga(\Ga\setminus\Ga')$ of the set of all smooth functions vanishing on
$\bar\Ga\cap\Ga_1$.
  This inequality together with \eq{3.74} with $\theta=3/4$
  yields
 $$\begin{array}{rcl}
 \int_{\Ga\setminus\Ga'}|d(x',\bar\Ga\cap\Ga_1)^{-1}v_\tau\cdot u_\tau|\,ds
 &\leq & \|d(x',\bar\Ga\cap\Ga_1)^{-3/4}u_\tau\|_{L^2(\Ga\setminus\Ga')}
  \|d(x',\bar\Ga\cap\Ga_1)^{-1/4}v_\tau\|_{L^2(\Ga\setminus\Ga')}\ek
 &\les &
 \|u_\tau\|_{H^{3/4}_{0,\bar\Ga\cap\Ga_1}(\Ga\setminus\Ga')}\|v_\tau\|_{H^{1/4}(\Ga\setminus\Ga')}
  \ek
&\les & \|u\|_{H^{3/4}(\Ga\setminus\Ga')}\|v\|_{H^{1/2}(\Ga)}
  \ek
& \les & \ve\|u\|_{W^{2,2}(\Om_0)}\|\na v\|_{L^2(\Om)}.
  \end{array}$$
Therefore, from \eq{3.68nn} it follows that
\begin{equation}
\label{E3.72}
\Big|\int_{\Ga\setminus\Ga'}v_\tau\cdot(\ve\Psi
 c^{bl})^{-1}u_\tau\,dx'\Big|\les \ve\|u\|_{W^{2,2}(\Om_0)}\|\na v\|_{L^2(\Om)}.
 \end{equation}
 Note that $\lan v_\tau,
\frac{\pa \bar\eta_\tau^\ve}{\pa\nu}\ran_{H^{\ha}(\Ga),
H^{-\ha}(\Ga)}= \lan \frac{\pa v_\tau}{\pa\nu},
\bar\eta_\tau^\ve\ran_{H^{-\ha}(\Ga), H^{\ha}(\Ga)}$ since
$v\cdot\nu|_{\Ga}=0, \eta\cdot\nu|_{\Ga}=0$. Therefore, by
\eq{3.70}, \eq{3.35nn} and \eq{3.4} we get that
\begin{equation}
\label{E3.73}
\begin{array}{rcl}
 |\lan v_\tau, \frac{\pa \bar\eta_\tau^\ve}{\pa\nu}\ran_{H^{\ha}(\Ga), H^{-\ha}(\Ga)}|
&=&|\lan \frac{\pa v_\tau}{\pa\nu},
\bar\eta_\tau^\ve\ran_{H^{-\ha}(\Ga), H^{\ha}(\Ga)}|\ek & = &|\lan
(\frac{\pa v}{\pa\nu})_\tau, \bar\eta_\tau^\ve\ran_{H^{-\ha}(\Ga),
H^{\ha}(\Ga)}|\ek
  &\leq &\ve\|\na v\|_2\|\eta\|_{H^1(\Om)}\ek
 &\les&
 \ve\|\na v\|_2(\|f\|_{L^q(\ome)}+\|\psi\|_{W^{2-1/q,q}(\Ga_1)}).
\end{array}
\end{equation}

 Thus, it follows from \eq{3.68n}, \eq{3.68nnn}, \eq{3.72} and \eq{3.73} that
\begin{equation}
\label{E3.69}
 \|\na v\|_2 \les
 \ve(\|f\|_{L^q(\ome)}+\|\psi\|_{W^{2-1/q,q}(\Ga_1)}),
 \end{equation}
which proves the first inequality of the theorem.

Next, let us prove the second inequality of the theorem.
Notice that \eq{3.62nn} holds for $v$ as well.
Then, we get from the boundary condition on $v$, uniform negativity of
matrix $c^{bl}(x')$ with respect to $x'\in \bar\Ga$, \eq{3.62nn}
with $v$ in place of $\vp$ and \eq{3.69n} that
$$\begin{array}{l}
\|v\|^2_{L^2(\pa\Om)}=\|v_\tau\|^2_{L^2(\Ga')}+\|v_\tau\|^2_{L^2(\Ga\setminus\Ga')}
+\|u_\nu\|^2_{L^2(\Ga\setminus\Ga')}\ek
 \les -(v_\tau,(c^{bl})^{-1}v_\tau)_{L^2(\Ga')}+\|v_\tau\|^2_{L^2(\Ga\setminus\Ga')}
  +\|u\|^2_{L^2(\Ga\setminus\Ga')}\ek
  \les -\ve\int_{\Ga'} v_\tau\cdot\frac{\pa}{\pa\nu}
 (v_\tau +\ve\eta_\tau)dx' + (v_\tau,(c^{bl})^{-1}u_\tau)_{L^2(\Ga\setminus\Ga')}
+\ve\|\na v\|^2_{L^2(\Om)} +\ve^3\|\na^2 u\|_{L^2(\Om_0)}\ek
  \les -\ve\int_{\Ga'} v_\tau\cdot\frac{\pa}{\pa\nu}
 (v_\tau +\ve\eta_\tau)dx' +\ve^2 \|\na v\|_{L^2(\Om)}\|\na^2 u\|_{L^2(\Om_0)}+\ve\|\na
v\|^2_{L^2(\Om)} +\ve^3\|\na^2 u\|_{L^2(\Om_0)}.
  \end{array}$$
In the right-hand side of this inequality, we get, in view of
\eq{3.70}, \eq{3.71} and \eq{3.4} that
$$\begin{array}{l}
 \ve|\int_{\Ga'} v_\tau\cdot\frac{\pa}{\pa\nu}
 (v_\tau +\ve\eta_\tau)dx'| = \ve|\int_{\Ga'} v_\tau\cdot(\frac{\pa}{\pa\nu}
 (v +\ve\eta))_\tau dx'|\ek
 \hspace{1cm}\les \ve\|v\|_{H^\ha(\Ga)}\|(\frac{\pa}{\pa\nu}
 (v +\ve\eta))_\tau \|_{H^{-\ha}(\Ga)}
  \ek
 \hspace{1cm}\les  \ve\|v\|_{H^1(\Ga_\da^\ve\cap\Om)}\|\na(v +\ve\eta)\|_{L^{2}(\Om)}
   \ek
 \hspace{1cm}\les \ve(\|\na v\|_{L^2(\Ga_\da^\ve\cap\Om)}
  +\|v\|_{L^2(\Ga_\da^\ve\cap\Om)})(\|\na v\|_{L^{2}(\Om)}
 +\ve\|\na\eta\|_{L^{2}(\Om)})
   \ek
  \hspace{1cm}\les \ve\|\na v\|^2_{L^2(\Om)}+\ve^2\|\na v\|_2\|\na\eta\|_{L^{2}(\Om)}
   +\ve^2\|v\|_{L^2(\Om)}(\|f\|_{L^q(\ome)}+\|\psi\|_{W^{1-1/q,q}(\Ga_1)}).
\end{array}$$
Therefore, by \eq{3.35nn}, \eq{3.4} and \eq{3.69} we have
\begin{equation}
\label{E3.80} \|v\|^2_{L^2(\pa\Om)}\les
\ve^3(\|f\|_{L^q(\ome)}+\|\psi\|_{W^{1-1/q,q}(\Ga_1)})^2
+\ve^2\|v\|_{L^2(\Om)}(\|f\|_{L^q(\ome)}+\|\psi\|_{W^{1-1/q,q}(\Ga_1)}).
\end{equation}
Thus, if we prove
\begin{equation}
\label{E3.79}\|v\|_{L^2(\Om)}
 \les \|v\|_{L^2(\pa\Om)}+\ve^{1/2}\|\na v\|_{L^2(\Om)},
 \end{equation}
then, in view of \eq{3.80} and the first inequality already
 proved, we have the second inequality of the theorem.

When $\Ga_1$ is a component of $\pa\Om$, \eq{3.79} is obvious from the property
 $\|v\|_{L^2(\Om)}  \les \|v\|_{L^2(\pa\Om)}$
 for a very weak solution to the Stokes system in $\Om$ of $C^2$-class (see e.g. \cite{GSS05}).
But, we can not claim $\|v\|_{L^2(\Om)}  \les \|v\|_{L^2(\pa\Om)}$
when $\Ga_1$ and $\Ga$ are adjacent, since $\Om$ is then $C^{0,1}$-domain. In
that case, let us choose a smooth subdomain $\Om'\subset\Om$
which is obtained by cutting off a very small
  tube $\Om\setminus\Om'$  from $\Om$
  such that
  $$\Om\setminus \Om'\subset \{x\in\Om: d(x,\Ga\cap \Ga_1)\leq \ve^{3/2}\}.$$
Then,  $|\Om\setminus\Om'|\sim \ve^3$. Note that, the estimate constant $C$ in the Sobolev
inequality $\|\vp\|_6\leq C\|\na \vp\|_2$ is invariant with respect
to scaling transforms, and hence we have
$\|v\|_{L^6(\Om\setminus\Om')}\leq c\|\na
v\|_{L^2(\Om\setminus\Om')}$ with constant $c$ independent of $\ve$.
Moreover, notice the inequality $\|v\|_{L^2(\Om')}\leq
c(\Om')\|v\|_{L^2(\pa\Om')}$  known for
very weak solutions to Stokes equations.
  Therefore, by Lemma \ref{L2.3} and \eq{3.80} we get that
 $$\begin{array}{rcl}
\|v\|^2_{L^2(\Om)} &\leq &
\|v\|^2_{L^2(\Om')}+\|v\|^2_{L^2(\Om\setminus\Om')}\ek
 &\les &
\|v\|^2_{L^2(\pa\Om')}+\|v\|^2_{L^6(\Om\setminus\Om')}\cdot|\Om\setminus\Om'|^{1/3}\ek
 &\les &\|v\|^2_{L^2(\pa\Om\cap\pa\Om')}
 +\|v\|^2_{L^2(\pa(\Om\setminus\Om')\cap\Om)}+\ve\|\na v\|^2_{L^2(\Om\setminus\Om')}.
  \end{array}$$
Here,  in the same way as the proof of
 Lemma \ref{L2.3}, we can get estimate
$$\|v\|^2_{L^2(\pa(\Om\setminus\Om')\cap\Om)}\les
\ve\|\na v\|^2_{L^2(\Om\setminus\Om')}+
\|v\|^2_{L^2(\pa\Om\setminus\pa\Om')}$$ using that
 $\Om$ is a Lipschitz domain.
 Consequently, we have
$$\begin{array}{rcl}
\|v\|^2_{L^2(\Om)}
 &\les &\|v\|^2_{L^2(\pa\Om\cap\pa\Om')}
 +\|v\|^2_{L^2(\pa\Om\setminus\pa\Om')}
  +\ve\|\na v\|^2_{L^2(\Om\setminus\Om')}\ek
  &= & \|v\|^2_{L^2(\pa\Om)}
  +\ve\|\na v\|^2_{L^2(\Om\setminus\Om')}
  \end{array}$$
and hence \eq{3.79}.

 The proof of the theorem is complete. \hfill $\Box$
\begin{lem}
\label{L3.18} {\rm Let $q>3$.
 For $\eta^\ve$ defined by \eq{3.38} there hold
\begin{equation}
\label{E3.64}
 \|\eta^\ve\|_{L^2(\Om)} \les
\ve^{3/2}(\|f\|_{L^q(\ome)}+\|\psi\|_{W^{2-1/q,q}(\Ga_1)})
\end{equation}
and
\begin{equation}
 \label{E3.65}
\|\na\eta^\ve\|_{L^1(\Om)} \les
\ve(\|f\|_{L^q(\ome)}+\|\psi\|_{W^{2-1/q,q}(\Ga_1)}).
\end{equation}
}
\end{lem}
{\bf Proof.} Due to embedding
$W^{1,q}(\Om)\subset L^\infty(\Om)$, \eq{3.44n}
and Lemma \ref{L3.12},  we have
$$\begin{array}{rcl}
\|\beta^{\ve,l}\chi_l\|_{L^2(\Om)}
 &\leq&
 \|\beta^{\ve,l}\|_{L^2(\Om)}\|\chi_l\|_{L^\infty(\Om)}\ek
 &\les & \|\beta^{\ve,l}\|_{L^2(\Om)}\|\chi_l\|_{W^{1,q}(\Om)}\ek
 &\les& \ve^{1/2}(\|f\|_{L^q(\ome)}+\|\psi\|_{W^{2-1/q,q}(\Ga_1)}),
 l=1\sim 3.
\end{array}$$
Hence, \eq{3.64} is proved in view of the construction of
$\eta^\ve$, see \eq{3.38}.

Note that $\|\na\beta^{\ve,l}\|_{L^1(\Om)}\les O(1)$ holds by
the construction of $\beta^{\ve,l}$, see \eq{3.35n}, and \eq{3.50} with $r=1$.
Hence it follows  that
$$\begin{array}{rcl}
\|\na(\beta^{\ve,l}\chi_l)\|_{L^1(\Om)}&\leq&
\|\na\beta^{\ve,l}\chi_l\|_{L^1(\Om)}+
 \|\beta^{\ve,l}\na\chi_l\|_{L^1(\Om)}\ek
&\leq& \|\na\beta^{\ve,l}\|_{L^1(\Om)}\|\chi_l\|_{L^\infty(\Om)}+
 \|\beta^{\ve,l}\|_{L^\infty(\Om)}\|\na\chi_l\|_{L^1(\Om)}\ek
 &\les& \|\chi_l\|_{L^\infty(\Om)}+\|\na\chi_l\|_{L^1(\Om)}\ek
 &\les&
 \|f\|_{L^q(\ome)}+\|\psi\|_{W^{2-1/q,q}(\Ga_1)},\,\,
 i,l=1\sim 3.
\end{array}$$
Thus, \eq{3.65} is proved.\vspace{0.2cm}
\hfill $\Box$
\par
By Theorem \ref{T3.15} and Lemma \ref{L3.18} we get the following
theorem showing the error estimate for the obtained
wall-law \eq{3.66}.
\begin{theo}
\label{T3.20} {\rm Let $f\in L^q(\ome),\psi\in W^{2-1/q,q}(\pa\ome),
q>3,$ and let $u^\ve$ and $u^{eff}$ be the solutions to the systems
\eq{3.1} and \eq{3.66}, respectively. Then,
 $$\begin{array}{lcl}
 \|u^\ve-u^{eff}\|_{L^2(\Om)} &\les&
\ve^{3/2}(\|f\|_{L^q(\ome)}+\|\psi\|_{W^{2-1/q,q}(\Ga_1)}), \ek
\|\na(u^\ve-u^{eff})\|_{L^1(\Om)} &\les&
\ve(\|f\|_{L^q(\ome)}+\|\psi\|_{W^{2-1/q,q}(\Ga_1)}).
\end{array}$$
}
\end{theo}
\begin{rem}
{\rm As seen above, the Navier-wall law derived in this work is independent of
the choice of the orthogonal tangent vectors; it depends only on the geometry
of the fictitious boundary $\Ga$ and roughness shape since
the matrix $c^{bl}$ in \eq{3.46} is constructed using boundary layers
near the rough surface, which are determined by the boundary layer cell problems $\PP$.

It will be shown in the forthcoming papers \cite{Ri12-1, Ri12-2}
that the results of boundary layer analysis
given in \S 3.2 are still fundamental for derivation of effective
wall-laws for Navier-Stokes equations over curved rough boundaries
as well as for fluid flows around rotating bodies.
For these problems, $c^{bl}$, constructed in \eq{3.46},
will also be shown to be useful coefficient matrix to be involved in the
effective wall-laws.
}
\end{rem}

\begin{rem}
{\rm
 The result of the paper can be directly extended to the case of
 spacial dimension $n>3$ without any nontrivial changes.
}
\end{rem}
\newpage

\begin{appendices}
\appendix
\section{Estimate for Divergence Problem $\div u=f$}

Divergence problem is one of the fundamental problems in the study of Navier-Stokes equations.
In some references rigorous estimates for some solutions of the divergence
problem is known, see e.g. \cite{Ga94-1}, Ch.III, Section 3.  Unfortunately,
however, the results of \cite{Ga94-1}
do not guarantee that for our domain $\Om^\ve$ given by \eq{1.2}
   the estimate constants for solutions to divergence equation
do not depend on the microscopic size $\ve$.
Therefore, in this appendix, we give a refined analysis for the dependence of the estimate
constant for solution to the divergence problem in some specific domains.\vspace{0.3cm}

\par
\noindent
{\bf Lemma A.1}
Let a simply connected and bounded domain $G$ of $R^n, n\geq 2,$
be expressed as
$$G=G_0\cup\bigcup_{k=1}^mG_k, \quad
G_0\cap G_k\neq \emptyset, \quad G_k\cap G_l=\emptyset (k\neq l), \quad  k,l=1,\ldots,m,$$
where $G_k, k=0,\ldots,m,$ has cone-property and star-shaped with respect to some balls $B(x_k,R_k)$
of radius $R_k$, and
 $\frac{{\rm diam}(G_0)}{R_0}+\frac{{\rm diam}(G_k)}{R_k}+\frac{|G_0|}{|G_0\cap G_k|}<l$
with some constant $l>0$ for all $k\in \{1,\ldots,m\}$.

If $f\in L^q(G)$, $1<q<\infty$, $\int_G f(x)\,dx=0$, then the divergence problem
\begin{equation}
\label{Ea1}
\div u=f\quad\text{in }G, \quad u|_{\pa G}=0,
\end{equation}
has a solution $u\in W^{1,q}_0(G)$ satisfying
\begin{equation}
\label{Ea2}
\|u\|_{W^{1,q}_0(G)}\leq C\|f\|_{L^q(G)}
\end{equation}
with constant $C=C(n,q,l)>0$ independent of $m$ and ${\rm diam}(G_k), k=0,\ldots,m$.
\vspace{0.2cm}
\par\noindent
{\bf Proof:}
Since the existence for the problem \eq{a1} is already well-known, see e.g. \cite{Ga94-1}, Ch.III, Theorem 3.1,
we shall show that the constant $C$ in \eq{a2} is irrespective
of $m$ and ${\rm diam}(G_k), k=0,\ldots,m,$ and depends only on $n,q$ and $l$.
For $k=1,\ldots,m$ let us define $f_k$ on $G_k$ by
$$f_k(x)=\left\{
        \begin{array}{cl}
         f(x)&\quad \text{for } x\in G_k\setminus G_0,\ek
         f(x)-a_k & \quad\text{for } x\in G_k\cap G_0,
        \end{array}
\right.
$$
where $a_k=\frac{\int_{G_k}f(x)\,dx}{|G_k\cap G_0|}$, and let $f_0:= f-\sum_{k=1}^m f_k$.
Obviously, $\supp f_k\subset G_k$, $\int_{G_k}f_k\,dx=0$
 for all $k\in \{0,\ldots,m\}$ and, denoting the extension
by $0$ of $f_k$ to $G$ again by $f_k$, we have $f=\sum_{k=0}^m f_k$.
Then, for $k=1,\ldots,m$  using H\"older inequality and
 $(a+b)^q\leq \bar{c}(q)(a^q+b^q)$ for $a,b\geq 0$
we get that
\begin{equation}
\label{Ea3}
\begin{array}{rcl}
\int_{G_k} |f_k|^q\, dx & = & \int_{G_k\setminus G_0} |f(x)|^q\,dx
         +\int_{G_k\cap G_0} |f(x)-a_k|^q\,dx\ek
      &\leq & \int_{G_k\setminus G_0} |f(x)|^q\,dx
       +\bar{c}(q)(\int_{G_k\cap G_0} |f(x)|^q\,dx+|a_k|^q|G_k\cap G_0|)\ek
      &=  & \bar{c}(q)(\int_{G_k} |f(x)|^q\,dx+|\int_{G_k}f(x)dx|^q|G_k\cap G_0|^{1-q})\ek
      &\leq& \bar{c}(q)\big(\int_{G_k} |f(x)|^q\,dx
        +\int_{G_k}|f(x)|^q\,dx \frac{|G_k|^{q-1}}{|G_k\cap G_0|^{q-1}}\big)\ek
      &\leq& \bar{c}(q)(1+l^{q-1})\int_{G_k} |f(x)|^q\,dx.
\end{array}
\end{equation}
Using \eq{a3} we get that
\begin{equation}
\label{Ea4}
\begin{array}{rcl}
\int_{G_0} |f_0(x)|^q\, dx & = & \int_{G_0\setminus \cup_{k=1}^m G_k} |f(x)|^q\,dx+
            \sum_{k=1}^m \int_{G_k\cap G_0} |f(x)+f_k|^q\,dx\ek
      &\leq & \int_{G_0\setminus \cup_{k=1}^m G_k} |f(x)|^q\,dx
         +\bar{c}(q)\sum_{k=1}^m(\int_{G_k\cap G_0} (|f(x)|^q+|f_k(x)|^q)\,dx)\ek
      &\leq & \int_{G_0\setminus \cup_{k=1}^m G_k} |f(x)|^q\,dx
         +\bar{c}(q)(1+\bar{c}(q)+l^{q-1})\sum_{k=1}^m\int_{G_k} |f(x)|^q\,dx\ek
       &=  & \bar{c}(q)(1+\bar{c}(q)+l^{q-1}) \int_{G} |f(x)|^q\,dx.
\end{array}
\end{equation}

On the other hand,  in view of the assumption $\frac{{\rm diam}(G_k)}{R_k}<l$, it
 follows by \cite{Ga94-1}, Ch.III, Theorem 3.1  that for $k=0,\ldots,m$
 the problem
$$\div u_k=f_k\quad\text{in }G_k,\quad u_k|_{\pa G_k}=0,$$
has a solution $u_k\in W^{1,q}_0(G_k)$ such that
\begin{equation}
\label{Ea5}
\|u_k\|_{W^{1,q}_0(G_k)}\leq c_0(n,q,l) \|f_k\|_{L^q(G_k)}
\end{equation}
with constant $c_0(n,q,l)$ independent of $k$. Thus, denoting extension by $0$ of $u_k$ to $G$
again by $u_k$, we get by cone-property of $G_k$ that $u:=\sum_{k=0}^m u_k\in W^{1,q}_0(\Om)$
and by \eq{a3}-\eq{a5} that
$$\begin{array}{rcl}
\|u\|^q_{W^{1,q}_0(G)}& = & \|u_0\|^q_{W^{1,q}_0(G_0\setminus \cup_{k=1}^m G_k)}
          + \sum_{k=1}^m \|u_0+u_k\|^q_{W^{1,q}(G_k)}\ek
      &\leq & \|u_0\|^q_{W^{1,q}_0(G_0\setminus \cup_{k=1}^m G_k)}
         +\bar{c}(q)\sum_{k=1}^m(\|u_0\|^q_{W^{1,q}(G_k\cap G_0)}+\|u_k\|^q_{W^{1,q}(G_k)})\ek
      &\leq & \bar{c}(q)\big(\|u_0\|^q_{W^{1,q}_0(G_0)}
         +\sum_{k=1}^m\|u_k\|^q_{W^{1,q}(G_k)}\big)\ek
      &\leq & \bar{c}(q)c_0(n,q,l)^q\big(\|f_0\|^q_{L^{q}(G_0)}
         +\sum_{k=1}^m\|f_k\|^q_{L^{q}(G_k)}\big)\ek
      &\leq & 2\bar{c}(q)c_0(n,q,l)^q(1+\bar{c}(q)+l^{q-1}) \|f\|^q_{L^q(G)}.
\end{array}
$$

Thus, \eq{a2} is proved with $C=c_0(n,q,l)(2\bar{c}(q)(1+\bar{c}(q)+l^{q-1}))^{1/q}$. \hfill $\Box$\vspace{0.3cm}
\par
In the next lemma we consider more general setting for the divergence problem.
\vspace{0.3cm}
\par\noindent
{\bf Lemma A.2}
Let a simply connected and bounded domain $G$ of $\R^n, n\geq 2,$
be expressed as
$$G=\bigcup_{j=1}^N G^{(j)}, \quad
G^{(j)}\cap G^{(j+1)}\neq \emptyset, \quad  j=1,\ldots,N-1,$$
where $G^{(j)}, j=1,\ldots,N,$ are simply connected domains with cone-property.
Moreover, suppose that  for each $j=1,\ldots,N$
the divergence problem \eq{a1} in $G^{(j)}$ has a solution $u\in
W^{1,q}_0(G^{(j)})$ satisfying \eq{a2} with constant $C_j>0$.
 If $f\in L^q(G)$, $1<q<\infty$, $\int_G f(x)\,dx=0$, then the divergence problem \eq{a1} in $G$
has a solution $u\in W^{1,q}_0(G)$ satisfying \eq{a2}
with constant $C>0$ bounded by linear combination of
$c(q)C_1\frac{(\min\{|G^{(1)}|, |G\setminus G^{(1)}|\})^{1-1/q}}{|G^{(1)}\cap G^{(2)}|^{1-1/q}}$
and
$c(q)C_j\frac{(\min\{|\cup_{i=1}^{j} G^{(i)}|, |\cup_{i=j+1}^{N}G^{(i)}|\})^{1-1/q}}{|(G^{(j)}\cap G^{(j+1)})\setminus \cup_{i=1}^{j-1}G^{(i)}|^{1-1/q}}$,
 $j=2,\ldots,N-1$.

\vspace{0.2cm}
\noindent
{\bf Proof:} First, construct functions $f_j\in L^q(G)$, $j=1,\ldots,N$, such that
\begin{equation}
\label{Ea8}
\supp f_j\subset G^{(j)},\quad \int_{G^{(j)}} f_j\,dx=0,\quad f(x)=\sum_{j=1}^N f^{(j)}(x),x\in G.
\end{equation}
Put
\begin{equation}
\label{Ea6}
f_1(x)=\left\{
        \begin{array}{cl}
         f(x)& \text{for } x\in G^{(1)}\setminus G^{(2)}\ek
         f(x)-a_1 & \text{for } x\in G^{(1)}\cap G^{(2)},
        \end{array}
\right.\quad
\quad
a_1=\frac{1}{|G^{(1)}\cap G^{(2)}|}\int_{G^{(1)}}f(x)\,dx
\end{equation}
and for $j=2,\ldots,N$
\begin{equation}
\label{Ea7}
f_j(x)=\left\{
        \begin{array}{cl}
         a_{j-1}   &\text{for } x\in (G^{(j)}\cap G^{(j-1)})\setminus \bigcup_{i=1}^{j-2} G^{(i)}\ek
         f(x)&\text{for } x\in G^{(j)}\setminus\big(G^{(j+1)}\cup \bigcup_{i=1}^{j-1} G^{(i)}\big)\ek
         f(x)-a_j &\text{for } x\in (G^{(j)}\cap G^{(j+1)})\setminus \bigcup_{i=1}^{j-1} G^{(i)}\ek
         0 &\text{for } x\in G^{(j)}\cap \bigcup_{i=1}^{j-2} G^{(i)}
        \end{array}
\right.
\end{equation}
with
$$a_j=\frac{\int_{\cup_{i=1}^j G^{(i)}}f(x)\,dx}
    {|(G^{(j)}\cap G^{(j+1)})\setminus \bigcup_{i=1}^{j-1} G^{(i)}|}.$$
Here and in what follows, $G^{(0)}=G^{(N+1)}\equiv \emptyset$ and hence
in \eq{a7} we neglect the cases where $G^{(0)}$ appears for $j=2$ or $G^{(N+1)}$ appears for $j=N$.

Denote the extension by $0$ of $f_j, j=1,\ldots,N,$ to $G$ again by $f_j$.
Then $f_j, j=1,\ldots,N,$ satisfy \eq{a8}.
In fact, it is clear that $\int_{G^{(1)}} f_1\,dx=0$, and for $j=2,\ldots, N$
 we have
 $$\begin{array}{l}
 \int_{G^{(j)}} f_j\,dx\ek
 =a_{j-1}|(G^{(j)}\cap G^{(j-1)})\setminus \bigcup_{i=1}^{j-2} G^{(i)}|
 -a_j |(G^{(j)}\cap G^{(j+1)})\setminus \bigcup_{i=1}^{j-1} G^{(i)}|
 + \int_{G^{(j)}\setminus \bigcup_{i=1}^{j-1} G^{(i)}}f(x)\,dx \ek
 =
  \int_{\bigcup_{i=1}^{j-1} G^{(i)}}f(x)\,dx
 -\int_{\bigcup_{i=1}^{j} G^{(i)}}f(x)\,dx
 + \int_{G^{(j)}\setminus \bigcup_{i=1}^{j-1} G^{(i)}}f(x)\,dx\ek
 =0.
  \end{array} $$
Moreover, $\sum_{j=1}^N f_j=f$ can be easily checked in view of the recursive construction of $f_j, j=1,\ldots,N$.

Now, let us get estimate of $\|f_j\|_{L^q(G^{(j)})}$, $j=1,\ldots,N$.
In view of $\int_B f(x)\,dx=\int_{G\setminus B} f(x)\,dx$ for any measurable set $B\subset G$,
 we get
 $$\begin{array}{rcl}
 \|f_1\|^q_{L^q(G^{(1)})}\,dx &\leq &
 \bar{c}(q)\big(\int_{G^{(1)}} |f(x)|^q\,dx
 +\frac{|\int_{G^{(1)}} f(x)\,dx|^q}{|G^{(1)}\cap G^{(2)}|^{q-1}}\big)\ek
& \leq &
 \bar{c}(q)\big(\int_{G^{(1)}} |f(x)|^q\,dx
 +\frac{\min\{\int_{G^{(1)}} |f|^q\,dx |G^{(1)}|^{q-1},\int_{G\setminus G^{(1)}} |f|^q\,dx |G\setminus G^{(1)}|^{q-1}\}}
 {|G^{(1)}\cap G^{(2)}|^{q-1}}\big)\ek
& \leq &
 \bar{c}(q)\big(1+\frac{\min\{|G^{(1)}|,|G\setminus G^{(1)}|\}^{q-1}}
 {|G^{(1)}\cap G^{(2)}|^{q-1}}\big)\|f\|^q_{L^q(G)}
    \end{array} $$
using the same technique of Lemma A.1. In the same way, for $j=2,\ldots,N-1$
we get that
 $$\begin{array}{l}
 \|f_j\|^q_{L^q(G^{(j)})}\,dx \ek
  \leq
 \bar{c}(q)\big(1+\frac{\min\{|\bigcup_{i=1}^{j-1} G^{(i)}|,|\bigcup_{i=j}^{N} G^{(i)}|\}^{q-1}}
 {|(G^{(j-1)}\cap G^{(j)})\setminus \bigcup_{i=1}^{j-2} G^{(i)}|^{q-1}}
    + \frac{\min\{|\bigcup_{i=1}^{j} G^{(i)}|,|\bigcup_{i=j+1}^{N} G^{(i)}|\}^{q-1}}
 {|(G^{(j)}\cap G^{(j+1)})\setminus \bigcup_{i=1}^{j-1} G^{(i)}|^{q-1}}\big)\|f\|^q_{L^q(G)}
    \end{array} $$
and for $j=N$
 $$ \|f_N\|^q_{L^q(G^{(N)})}\,dx   \leq
 \bar{c}(q)\big(1+\frac{\min\{|\bigcup_{i=1}^{N-1} G^{(i)}|,|G^{(N)}|\}^{q-1}}
 {|(G^{(N-1)}\cap G^{(N)})\setminus \bigcup_{i=1}^{N-2} G^{(i)}|^{q-1}}
    \big)\|f\|^q_{L^q(G)}.
$$

Thus, by the assumption of the lemma, for each $j=1,\ldots, N$
the divergence problem \eq{a1} in $G^{(j)}$ with $f_j$ in the right-hand side
has a solution $u_j\in W^{1,q}_0(G^{(j)})$
such that
$$\|u_j\|_{W^{1,q}_0(G^{(j)})}\leq C_j \tilde{C}_j$$
where
$$\tilde{C}_j\leq \bar{c}(q)^{1/q}
 \Big(1+\Big(\frac{\min\{|\bigcup_{i=1}^{j-1} G^{(i)}|,|\bigcup_{i=j}^{N} G^{(i)}|\}}
 {|(G^{(j-1)}\cap G^{(j)})\setminus \bigcup_{i=1}^{j-2} G^{(i)}|}\Big)^{1-1/q}
    + \Big(\frac{\min\{|\bigcup_{i=1}^{j} G^{(i)}|,|\bigcup_{i=j+1}^{N} G^{(i)}|\}}
 {|(G^{(j)}\cap G^{(j+1)})\setminus \bigcup_{i=1}^{j-1} G^{(i)}|}\Big)^{1-1/q}\Big);$$
when $j=1$ the second term in the bracket of the right-hand side is neglected and
when $j=N$ the third term is neglected.

Obviously, $u=\sum_{j=1}^N u_j$ solves \eq{a1} with right-hand side $f$ and
the estimate \eq{a2} holds with constant $C$ bounded by a sum of
$\bar{c}(q)C_l\frac{(\min\{|\bigcup_{i=1}^{j-1} G^{(i)}|,|\bigcup_{i=j}^{N} G^{(i)}|\})^{1-1/q}}
 {|(G^{(j-1)}\cap G^{(j)})\setminus \bigcup_{i=1}^{j-2} G^{(i)}|^{1-1/q}}$.
 \hfill $\Box$
\end{appendices}

%
%
%


\end{document}